\documentclass[preprint]{iucr}
\usepackage{latexsym,amssymb,amsfonts,amsmath,amscd}
\allowdisplaybreaks
\usepackage{graphicx}
\usepackage{subfigure}

\usepackage{color}

\newcommand{\nouno}{\par\vspace{1mm}\noindent}

\begin{document}

\title{X-ray and $\gamma$-ray propagation in bent crystals with flat and cylindrical surfaces}
\shorttitle{X-ray and $\gamma$-ray propagation in bent crystals}

\cauthor[a,b]{A.}{Apolloni}{apolloni@to.infn.it}{}
\author[c]{G.}{Mana}{}{}
\author[d,a]{C.}{Palmisano}{}{}
\author[a]{G.}{Zosi}{}{}
\aff[a]{Universit\`a di Torino,
Dipartimento di Fisica Generale ``A.~Avogadro", via~P.~Giuria~1, \city{Torino},
\country{Italy}}
\aff[b]{NDSSL-Network Dynamics and Simulation Science Laboratory,
Virginia Bioinformatics Institute,
Virginia Polytechnic Institute and State University,
Research Building XV (0477) 1880 Pratt Drive, \city{Blacksburg}, \country{USA}}
\aff[c]{INRiM-Istituto Nazionale di Ricerca  Metrologica,
str.~delle Cacce~91, \city{Torino},
\country{Italy}}
\aff[d]{Museo Storico della Fisica e Centro Studi e
Ricerche ``Enrico~Fermi'', via~Panisperna~89A, \city{Roma}, \country{Italy}}

\nouno
\begin{abstract}

In this paper we investigate x-ray and $\gamma$-ray propagation 
in crystals having a constant strain gradient and flat or cylindrical surfaces.
When a displacement field is present, we solve the Takagi-Taupin equations
either by  the Riemann-Green method or by a  numerical method.
We apply the results  to study the operation of a double-crystal Laue-Laue
diffractometer having a flat collimating crystal followed by a bent analyzer crystal.
In particular, we examine the effect of the analyzer strain on the 
location of the diffraction peaks in the dispersive and non-dispersive setup,
thus confirming our previously reported peak-location as being  set only by the
diffracting plane spacing on the analyzer entrance surface.
\end{abstract}
\maketitle
%
\section{Introduction}
The main purpose of this work is the to verify the soundness
of what we had previously published (Mana \emph{et al.}, 2004).
On that occasion, we had studied how a constant strain gradient in the
rotating  crystal of a double-crystal diffractometer affects the instrument operation.
The result of that investigation was that the
position of the Bragg peaks depends only on the diffracting-plane spacing on the
crystal entrance-surface. However, to simplify the problem, we had then assumed all
the crystal surfaces rigorously flat.
Since subsequent experiments,
using cylindrically bent crystals and both x- and $\gamma$-rays
to test that rather surprising result, 
delivered contradictory results still under examination (Kessler 2007, Massa \emph{et al.}, 2005),
we were urged to investigate propagation in bent crystals
in more detail. An additional reason was the interest in efficient
Laue-Laue bent crystal diffractometers for $\gamma$-ray spectroscopy of
nuclei having a very high thermal neutron cross-section (Materna \emph{et al.}, 2006).

For these reasons, we extend here our previous analysis by taking  account of 
the curvature of the crystal surfaces and by simulating diffraction  
in both the non-dispersive and dispersive setups.
In section \ref{t-t} we solve, by the Riemann-Green method in cartesian coordinates,
the Takagi-Taupin equations for the propagation of x- and $\gamma$-rays in bent crystals;
we indicate also how the crystal surfaces are modelled, what choice of the
reference perfect-crystal we have adopted and what kinds of distortion 
of the cylindrically bent crystal we have  considered.
Sections \ref{flat} and \ref{curved} deal with two cases when the crystal surfaces,
on which the initial conditions have to be assigned,
are flat or cylindrical, respectively. While in the first case the solutions are known
(Authier \& Simon 1968, Mana \& Palmisano  2004),
and  are re-examined here to illustrate our formalism, we are not aware
of solutions when the crystal surface is a cylinder.
In fact, in the literature, 
the  case of a curved crystal surface in the macroscopic sense is
just hinted in Takagi (1969) by means of a curvilinear  coordinates.
Subsequently, Olekhnovitch \& Olekhnovitch (1980) carried out the calculation of the
profile function of the scattering curve for a crystal
in the form of a cylinder the  size of which does not exceed the extinction length.
Later, Thorkildsen \& Larsen (1998) observed that
it is in principle possible to obtain analytical expressions for the
primary extinction factor in perfect crystals having a circular diffraction plane.
In section \ref{double} we apply our results to a double diffractometer and,
to corroborate them, in section \ref{num-sim} we solve  numerically the Takagi-Taupin equations
in polar coordinates  for different
geometrical and physical parameters.

\newpage
\section{Takagi-Taupin equations for distorted crystals}\label{t-t}
In order to study x-ray propagation through a distorted crystal,
we shall apply the Takagi-Taupin equations
(Takagi 1962, Takagi 1969, Taupin 1964, Autier 2001, Mana \& Montanari 2004)
in the two-wave approximation of the dynamical theory of x-ray diffraction.
Since only cylindrical geometries will be considered, we shall use a two-dimension
model with a reference frame having the $x$ and $z$ axes
lying in the reflection plane.
Lattice distortion is described by the displacement field $\mathbf{u}(x,z)$,
which gives the difference between the actual distorted lattice and a virtual
perfect-lattice identified by the reciprocal vector $\mathbf{h}_0$,
which will be chosen according to our convenience.
Hence, we anchor the reference frame to it and set
the $x$ axis parallel or antiparallel to $\mathbf{h}_0$. By using directional derivatives,
the Takagi-Taupin equations can be written as

\begin{subequations}
\begin{eqnarray}
 -\mathrm{i} \frac{\partial D_\mathrm{o}}{\partial \mathbf{\hat{s}}_\mathrm{o}}
 &=&
 \frac{K \chi_\mathrm{o}}{2} D_\mathrm{o}+
 \frac{K \chi_{-\mathrm{h}}}{2} D_\mathrm{h}
 \label{TDor} \\
 -\mathrm{i} \frac{\partial D_\mathrm{h}}{\partial \mathbf{\hat{s}}_\mathrm{h}}
 &=&
 \frac{K \chi_\mathrm{o}}{2} D_\mathrm{h} +
 \frac{K \chi_\mathrm{h}}{2} D_\mathrm{o} +
 \frac{\partial (\mathbf{h}_0 \cdot \mathbf{u})}{\partial \mathbf{\hat{s}}_\mathrm{h}} D_\mathrm{h}.
 \label{TDhr}
\end{eqnarray}
\end{subequations}
In equations (\ref{TDor}) and (\ref{TDhr}), $D_\mathrm{o}$ and $D_\mathrm{h}$ are
slowly varying amplitudes of the Ewald expansion

\begin{equation}
 D = D_\mathrm{o} \exp{\left(\mathrm{i} \mathbf{K}_\mathrm{o} \cdot \mathbf{r}\right)}
 + D_\mathrm{h} \exp{\left[\mathrm{i} (\mathbf{K}_\mathrm{h}
  \cdot \mathbf{r}-\mathbf{h}_0 \cdot \mathbf{u})\right]}
\label{decomp}
\end{equation}
of the dielectric displacement vector $\mathbf{D}= D \mathbf{\hat{y}}$ for the $\sigma$ polarization,
\begin{subequations}
\begin{eqnarray}
\mathbf{\hat{s}}_\mathrm{o}&=& \mathbf{\hat{K}}_\mathrm{o} = -{\bf \hat{x}}\sin{\theta_\mathrm{B}} +
 {\bf \hat{z}}\cos{\theta_\mathrm{B}}
\label{veccar}\\
\mathbf{\hat{s}}_\mathrm{h}&=& \mathbf{\hat{K}}_\mathrm{h} = {\bf \hat{x}}\sin{\theta_\mathrm{B}} +
 {\bf \hat{z}}\cos{\theta_\mathrm{B}}
\label{veccarh}
\end{eqnarray}
\end{subequations}
are the unit propagation vectors, $\mathbf{h}_0=\mathbf{\hat{x}}2 K\sin{\theta_\mathrm{B}}$,
 $\mathbf{K}_\mathrm{h}=\mathbf{K}_\mathrm{o}+\mathbf{h}_0$, $K$
$=\|\mathbf{K}_\mathrm{o}\|=\|\mathbf{K}_\mathrm{h}\|=2\pi\nu/c$
is the modulus of the wave number vector of the incoming radiation
(with frequency $\nu$), $\theta_\mathrm{B}$ is the Bragg angle (with a sign),
and the complex parameters
$\chi_\mathrm{o}$, $\chi_{\mathrm{h}}$, and $\chi_{-\mathrm{h}}$ are the
Fourier components of electric susceptibility; in our case
$\chi_{-\mathrm{h}}=\chi_{\mathrm{h}}$.
We shall consider an infinite crystal slab the surfaces of which are smooth curves,
$\Gamma:[\tau_1,\,\tau_2] \rightarrow \mathbf{R^2}$ and
$\Sigma:[\sigma_1,\,\sigma_2] \rightarrow \mathbf{R^2}$,
with $\|\mathrm{d}\Gamma/\mathrm{d}\tau\|$ and $\|\mathrm{d}\Sigma/\mathrm{d}\sigma\ \|\in ]0,1/\tan\theta_\mathrm{B}[$,
where the last constraint is necessary to have a well-posed problem.
By imposing that at each point of the entrance surface the o-component of
the Ewald expansion (\ref{decomp}) is equal to the incoming
wave $D_\mathrm{e}(x,z)=A(x,z)\exp(\mathrm{i}\mathbf{K_\mathrm{e}}\mathbf{r})$,
the initial conditions to solve equations (\ref{TDor}, \ref{TDhr}) are therefore
\begin{subequations}
\begin{eqnarray}
\left.D_\mathrm{o}(x,z)\right|_{\Gamma}&=&\left.\Phi(x,z)\right|_{\Gamma}\\
\left.D_\mathrm{h}(x,z)\right|_{\Gamma}&=&0\ ,
\label{boundD}
\end{eqnarray}
\end{subequations}
where
$\Phi(x,z)=A(x,z)\exp\big[\mathrm{i}(\mathbf{K}_\mathrm{e}-\mathbf{K}_\mathrm{o})\mathbf{r}\big]$.

Equations (\ref{TDor}, \ref{TDhr}) can be simplified by introducing
two new amplitudes, $d_\mathrm{o}$ and $d_\mathrm{h}$, defined as
\begin{equation}
d_\mathrm{o,h}=
\exp
\left(
-\mathrm{i}\frac{K\chi_\mathrm{o}}{2}
\frac{\mathbf{\hat{s}}_\mathrm{o}+
\mathbf{\hat{s}}_\mathrm{h}}{1+\mathbf{\hat{s}}_\mathrm{o}\cdot\mathbf{\hat{s}}_\mathrm{h}}
\cdot\mathbf{r}
\right)
D_\mathrm{o,h}.
\label{subr}
\end{equation}
By substituting equations (\ref{subr}) for $D_\mathrm{o,h}$ into equations
(\ref{TDor}, \ref{TDhr}), we obtain

\begin{subequations}
\begin{eqnarray}
 -\mathrm{i} \frac{\partial d_\mathrm{o}}{\partial \mathbf{\hat{s}}_\mathrm{o}}
 &=&
 \frac{K \chi_{-\mathrm{h}}}{2} d_\mathrm{h} \label{TDo}\\
 -\mathrm{i} \frac{\partial d_\mathrm{h}}{\partial \mathbf{\hat{s}}_\mathrm{h}}
 &=&
 \frac{K \chi_{\mathrm{h}}}{2} d_\mathrm{o}
 +
 \frac{\partial (\mathbf{h}_0 \cdot \mathbf{u})}{\partial
  \mathbf{\hat{s}}_\mathrm{h}} d_\mathrm{h}, \label{TDh}
\end{eqnarray}
\end{subequations}
%
with the initial conditions
%
\begin{subequations}
\begin{eqnarray}
\left.d_\mathrm{o}(x,z)\right|_{\Gamma}&=&
\left.
\exp
\left(
-\mathrm{i}\frac{K\chi_\mathrm{o}}{2}
\frac{\mathbf{\hat{s}}_\mathrm{o}+\mathbf{\hat{s}}_\mathrm{h}}{1+
\mathbf{\hat{s}}_\mathrm{o}\cdot\mathbf{\hat{s}}_\mathrm{h}}
\cdot\mathbf{r}
\right)
\Phi(x,z)
\right|_{\Gamma}
\label{boundDo}\\
\left.d_\mathrm{h}(x,z)\right|_{\Gamma}&=&0\ .
\label{boundDh}
\end{eqnarray}
\end{subequations}
%
As in the cylindrical crystals the geometry is somewhat elaborate,
in the following subsections we illustrate the main underlying assumptions.
\subsection{Crystal surfaces} \label{cry-surf}
Let us now exemplify how cylindrical surfaces are modelled.
The curvature centre can be located either on the source side or on the opposite,
with $R_0$ and $R_0+T$ ($T$ denoting the crystal thickness) being the
curvature radii of the entrance surface, respectively.
A crystals having cylindrical surfaces will be called concave when its concavity
is towards the source, convex otherwise.

In the concave case, by locating the reference-frame origin at the curvature centre,
the parametric components of the surfaces are
\begin{equation}
 \left\{
 \begin{array}{ll}
 \Gamma_{x}(\tau)=R_0\sin\tau\\
 \Gamma_{z}(\tau)=R_0\cos\tau
 \end{array} \right.
 \label{GammaR}
\end{equation}
and
\begin{equation}
 \left\{
 \begin{array}{ll}
 \Sigma_{x}(\sigma)=(R_0+T)\sin\sigma\\
 \Sigma_{z}(\sigma)=(R_0+T)\cos\sigma
 \end{array} \right. ,
 \label{SigmaR}
\end{equation}
where $\tau \in [\tau_1,\;\tau_2]$ and $\sigma\in[\sigma_1,\;\sigma_2]$.
The inward and outward normals to the entrance and exit surfaces, respectively, are
\begin{equation}
 \left\{
 \begin{array}{ll}
 (\mathbf{\hat{n}}_\Gamma)_{x}(\tau)=\sin\tau\\
 (\mathbf{\hat{n}}_\Gamma)_{z}(\tau)=\cos\tau
 \end{array} \right.
 \label{Gamman}
\end{equation}
and
\begin{equation}
 \left\{
 \begin{array}{ll}
 (\mathbf{\hat{n}}_\Sigma)_{x}(\sigma)=\sin\sigma\\
 (\mathbf{\hat{n}}_\Sigma)_{z}(\sigma)=\cos\sigma
 \end{array} \right. .
 \label{Sigman}
\end{equation}
In the convex case, by locating the reference-frame origin again at the
curvature centre, the surface components and their inward and outward normals are
\begin{equation}
 \left\{
 \begin{array}{ll}
 \Gamma_{x}(\tau)=(R_0+T)\sin\tau\\
 \Gamma_{z}(\tau)=(R_0+T)\cos\tau
 \end{array} \right. ,
 \label{GammaRT}
\end{equation}

\begin{equation}
 \left\{
 \begin{array}{ll}
 \Sigma_{x}(\sigma)=R_0\sin\sigma\\
 \Sigma_{z}(\sigma)=R_0\cos\sigma
 \end{array} \right. ,
 \label{SigmaRT}
\end{equation}

\begin{equation}
 \left\{
 \begin{array}{ll}
 (\mathbf{\hat{n}}_\Gamma)_{x}(\tau)=-\sin\tau\\
 (\mathbf{\hat{n}}_\Gamma)_{z}(\tau)=-\cos\tau
 \end{array} \right. ,
 \label{GammanT}
\end{equation}
and
\begin{equation}
 \left\{
 \begin{array}{ll}
 (\mathbf{\hat{n}}_\Sigma)_{x}(\sigma)=-\sin\sigma\\
 (\mathbf{\hat{n}}_\Sigma)_{z}(\sigma)=-\cos\sigma
 \end{array} \right. .
 \label{SigmanT}
\end{equation}
%
\subsection{Crystal rotation} \label{cry-rot}
With a rotated crystal, since the reference perfect-lattice can be chosen at our convenience,
we make the virtual lattice immovable, no matter what the crystal rotation might be;
consequently, a rotation is nothing that a very special kind of distortion.
Hence, our {\it Ans\"atze} to study x-ray propagation in a rotated crystal are: firstly, $\mathbf{h}_0$ is
independent of rotations, secondly, the first-order rotation term
\begin{equation}
 \mathbf{u}^\mathrm{rot}(x,z) = [z\sin\alpha + x(\cos\alpha-1)] {\bf \hat{x}} \approx \alpha z {\bf \hat{x}},
 \label{urotcar}
\end{equation}
where $\alpha$ is the rotation angle (clockwise oriented),
must be added to the crystal distortion and, thirdly,
the crystal surfaces must be mapped into the new lines
$(\Gamma_x',\Gamma_z')=(\Gamma_x\cos\alpha+\Gamma_z\sin\alpha,-\Gamma_x\sin\alpha+\Gamma_z\cos\alpha)$
and
$(\Sigma_x',\Sigma_z')=(\Sigma_x\cos\alpha+\Sigma_z\sin\alpha,-\Sigma_x\sin\alpha+\Sigma_z\cos\alpha)$.
As long as $\alpha\ll 1$, we shall consider $\cos\alpha\approx 1$ and $\sin\alpha \approx 0$,
so that the crystal surfaces are left unchanged by rotations.
\subsection{Lattice distortion} \label{lat-def}
\begin{figure}
\begin{minipage}[t]{0.45\textwidth}
\begin{center}
\includegraphics[width=6.0 cm]{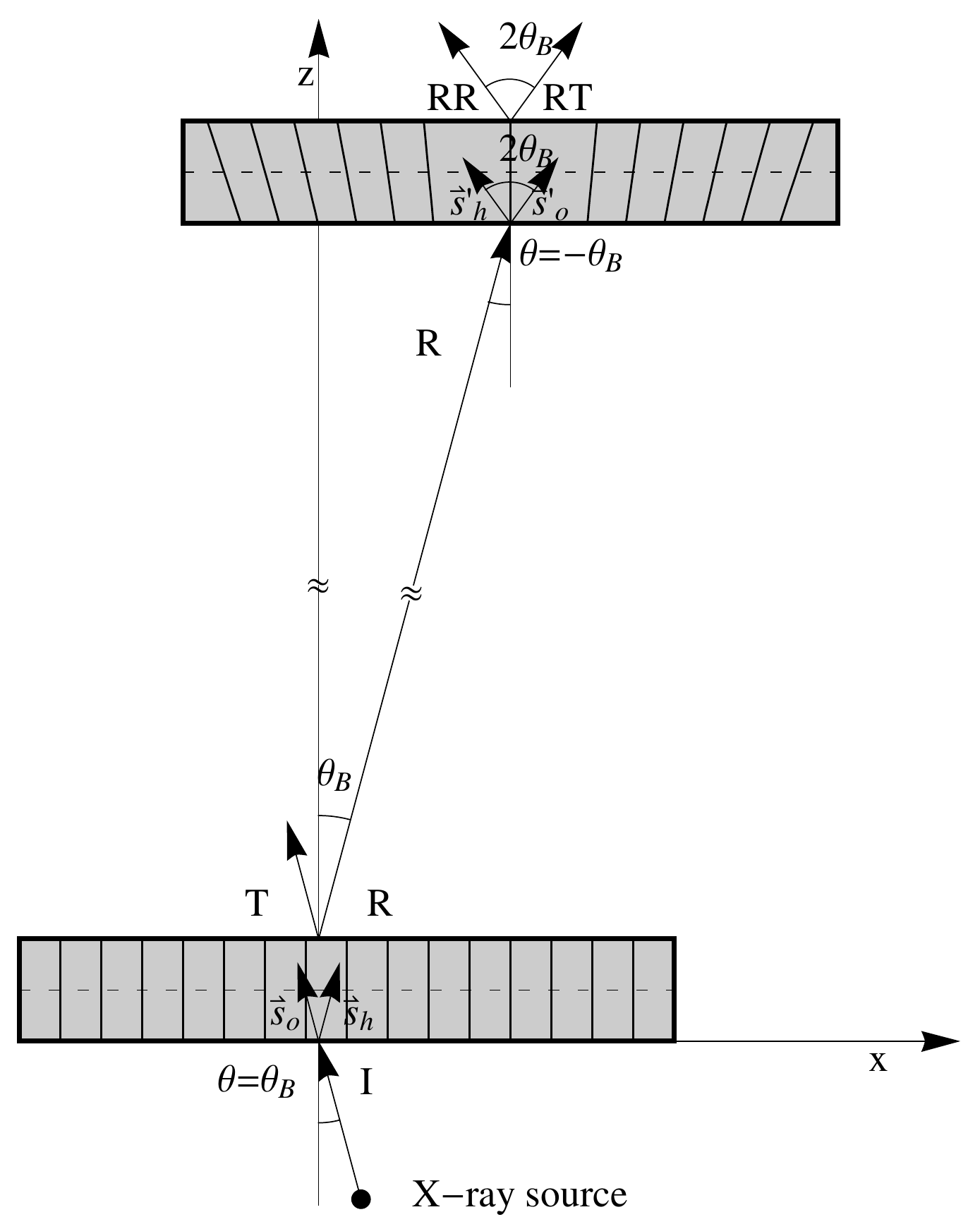}
\end{center}
\end{minipage}
\hfill
\begin{minipage}[t]{0.45\textwidth}
\begin{center}
\includegraphics[width=6.0 cm]{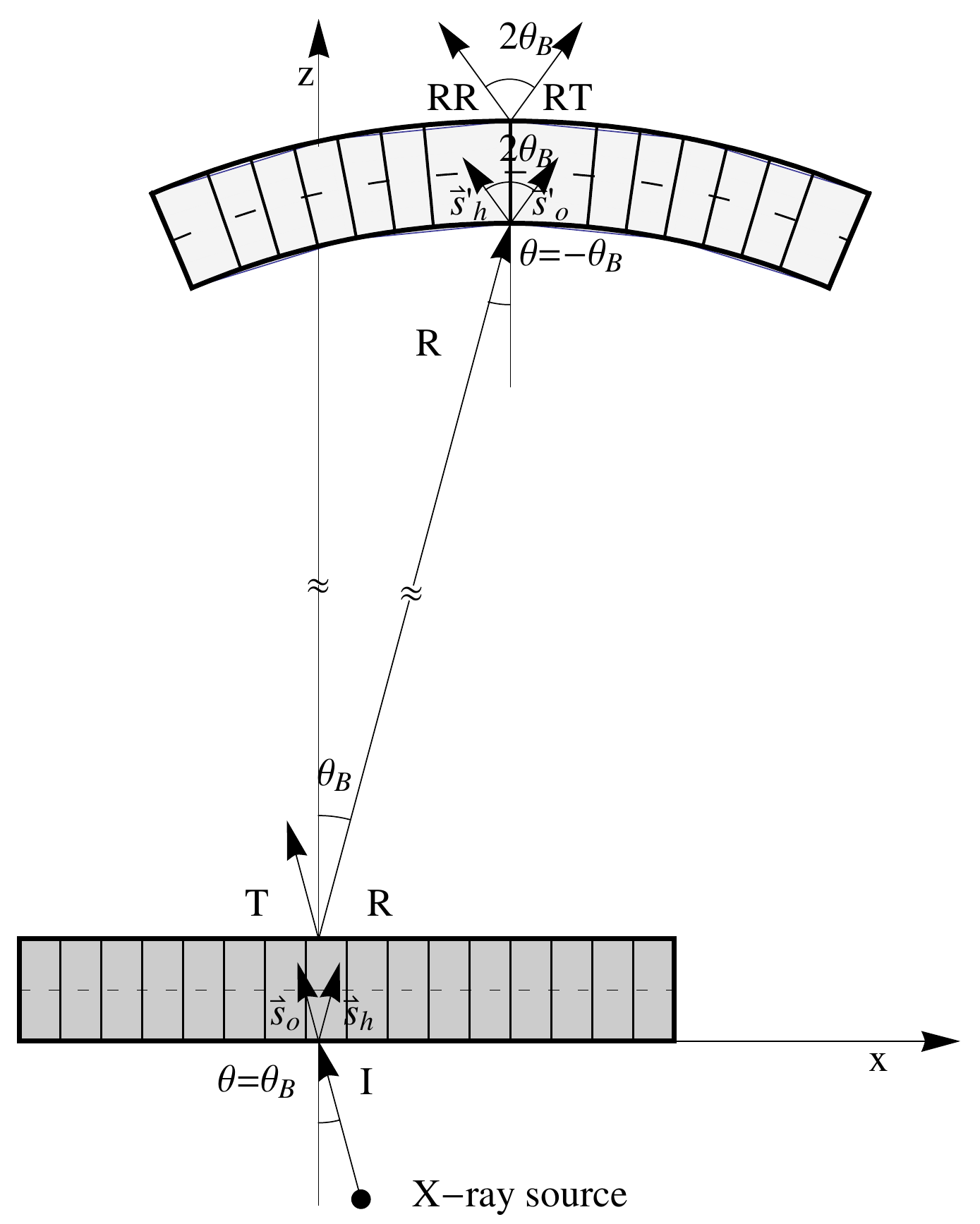}
\end{center}
\end{minipage}
\caption{
Layout of a two-crystal diffractometer in non-dispersive geometry.
On the left,
the analyzer crystal surface is flat and on the right it is cylindrically bent.
The symbol I is the incident beam, T and R are the beams transmitted and reflected
by the collimating crystal, RT and RR are the beams transmitted and reflected
by the analyzer crystal. The unit vectors ${\bf \hat{s}_{o,h}}$ are
defined in equations (\ref{veccar}) and (\ref{veccarh}). 
The angle of incidence $\theta$ is positive 
($\theta=\theta_\mathrm{B}$) on the collimating crystal
and negative ($\theta=-\theta_\mathrm{B}$) on the analyzer crystal.
}
\label{f1}
\end{figure}

We shall consider two distortions describing cylindrically bent crystals; both, fan-down,
\begin{subequations}
\begin{equation}
 v_{x}^{(1)}(x,z) = \frac{x(z-z_\mathrm{m})}{R_0+z_\mathrm{m}} ,
 \label{ucard}
\end{equation}
and fan-up,
\begin{equation}
 v_{x}^{(2)}(x,z) = \frac{x(z_\mathrm{m}-z)}{R_0+z_\mathrm{m}} ,
 \label{ucaru}
\end{equation}
\end{subequations}
are pure displacements in the $x$ direction.
We have chosen the $x$ axis origin and the
reference perfect-lattice so that, for the
non-rotated crystal, $\mathbf{u}(x=0, z_\mathrm{m})=0$ and $\mathbf{h}_0=\mathbf{h}(x,z=z_\mathrm{m})$,
where $\mathbf{h}$ is the local reciprocal vector of the distorted lattice.
In equation (\ref{ucard}), the displacement is such that all the lattice planes
are directed towards a point at distance $R_0$ from the entrance surface and the points
at $z=z_\mathrm{m}$ are undisplaced (left upper part of Fig.~\ref{f1}).
In equation (\ref{ucaru}), the lattice planes are directed towards a point at distance $R_0$
from the exit surface.
For a rotated crystal, the complete displacement field, including both equations
(\ref{ucard}, \ref{ucaru}) and equation (\ref{urotcar}), is
\begin{eqnarray}
 u_{x}^{(i)}(x,z)
 &=&
 v_{x}^{(i)}(x,z)+u_x^\mathrm{rot}(x,z)\nonumber\\
 &=&
 g(i) \frac{x(z-z_\mathrm{m})}{R_0+z_\mathrm{m}} + \alpha z\,,
\label{contr}
\end{eqnarray}
where $g(1)=1$, $g(2)=-1$
and the approximation $v_{x}^{(i)}(x,z)\cos\alpha\approx v_{x}^{(i)}(x,z)$ is applied.
\section{Propagation in distorted crystals}
Equations (\ref{ucard}) and (\ref{ucaru}) approximate the displacement field, characterized by a
constant strain gradient, in crystals having their surfaces flat or cylindrical.
\subsection{Flat crystal surfaces}\label{flat}
In the simplest case of flat external crystal surfaces, the Takagi-Taupin equations are
\begin{subequations}
\begin{eqnarray}
&-&\sin{\theta_\mathrm{B}}\frac{\partial d_\mathrm{o}^{(i)}}{\partial x} +
\cos{\theta_\mathrm{B}}\frac{\partial d_\mathrm{o}^{(i)}}{\partial z}
=
\mathrm{i}\frac{K\chi_{-\mathrm{h}}}{2} d_\mathrm{h}^{(i)}
\label{TDocar}\\
&&\sin{\theta_\mathrm{B}} \frac{\partial d_\mathrm{h}^{(i)}}{\partial x} +
\cos{\theta_\mathrm{B}}\frac{\partial d_\mathrm{h}^{(i)}}{\partial z}
=
\mathrm{i}\frac{K\chi_{\mathrm{h}}}{2} d_\mathrm{o}^{(i)}
\nonumber\\
&+&
\mathrm{i}2 K\sin{\theta_\mathrm{B}}
\left[
g(i)\frac{(z-z_\mathrm{m})}{R_0+z_\mathrm{m}}\sin{\theta_\mathrm{B}}
+
\left(g(i)\frac{x}{R_0+z_\mathrm{m}}+\alpha\right)\cos{\theta_\mathrm{B}}
\right]
d_\mathrm{h}^{(i)}
\label{TDhcar}
\end{eqnarray}
\end{subequations}
with the initial conditions
\begin{subequations}
\begin{eqnarray}
d_\mathrm{o}^{(i)}(x,0)&=&\Phi(x)\label{ico}\\
d_\mathrm{h}^{(i)}(x,0)&=&0, \label{ich}
\end{eqnarray}
\end{subequations}
where we assumed that the  external crystal surfaces
are orthogonal to $\mathbf{\hat{s}}_\mathrm{o}+\mathbf{\hat{s}}_\mathrm{h}$
(symmetrical Laue geometry)
and we located $z=0$ on the entrance surface.

The coupled equations (\ref{TDocar}, \ref{TDhcar}) can be simplified
by a change of dependent variables.
Let us introduce the two unknowns
$\tilde{D}_\mathrm{o}^{(i)}$ and $\tilde{D}_\mathrm{h}^{(i)}$ defined by the expression
\begin{equation}
\tilde{D}_\mathrm{o,h}^{(i)}
=
\exp(-\mathrm{i}f^{(i)} 2 K\sin\theta_\mathrm{B})d_\mathrm{o,h}^{(i)} \,,
\label{sub}
\end{equation}
where the function $f^{(i)}(x,z)$ is defined as
\begin{eqnarray}
f^{(i)}(x,z)
&=&
g(i)
\frac{1}{R_0+z_m}
\left[
\frac{1}{4}
\left(
\frac{x}{\sin\theta_\mathrm{B}}+\frac{z}{\cos\theta_\mathrm{B}}
\right)^2
\sin\theta_\mathrm{B}\cos\theta_\mathrm{B}-
z_m
\frac{1}{2}
\left(
\frac{x}{\sin\theta_\mathrm{B}}+\frac{z}{\cos\theta_\mathrm{B}}
\right)
\sin\theta_\mathrm{B}\right]\nonumber\\
&+&
\alpha
\frac{1}{2}
\left(
\frac{x}{\sin\theta_\mathrm{B}}+\frac{z}{\cos\theta_\mathrm{B}}
\right)\cos\theta_\mathrm{B}\,.
\label{fi}
\end{eqnarray}
If we observe that
\begin{eqnarray}
&&\left(\sin\theta_\mathrm{B}\frac{\partial}{\partial x}+\cos{\theta_\mathrm{B}}\frac{\partial}{\partial z}\right)f^{(i)}(x,z)
=\nonumber\\
&&
\sin{\theta_\mathrm{B}}\ g(i)\frac{(z-z_\mathrm{m})}{R_0+z_\mathrm{m}}
+
\cos{\theta_\mathrm{B}}\left(g(i)\frac{x}{R_0+z_\mathrm{m}}+\alpha\right)
\end{eqnarray}
and
\begin{equation}
\left(-\sin{\theta_\mathrm{B}}\frac{\partial}{\partial x}+\cos{\theta_\mathrm{B}}\frac{\partial}{\partial z}\right)f^{(i)}(x,z)
=
0\,,
\end{equation}
substitution of equation (\ref{sub}) into equations (\ref{TDocar}, \ref{TDhcar})
gives the Takagi-Taupin equations in the unperturbed form
\begin{subequations}
\begin{eqnarray}
&-&\sin{\theta_\mathrm{B}}\frac{\partial\tilde{D}_\mathrm{o}^{(i)} }{\partial x}+
\cos{\theta_\mathrm{B}}\frac{\partial\tilde{D}_\mathrm{o}^{(i)} }{\partial z}
=
\mathrm{i}\frac{K\chi_{-\mathrm{h}}}{2}\tilde{D}_\mathrm{h}^{(i)}
\label{TDotilde}\\
& &\sin{\theta_\mathrm{B}}\frac{\partial\tilde{D}_\mathrm{h}^{(i)} }{\partial x}+
\cos{\theta_\mathrm{B}}\frac{\partial\tilde{D}_\mathrm{h}^{(i)} }{\partial z}
=
\mathrm{i}\frac{K\chi_{\mathrm{h}}}{2}\tilde{D}_\mathrm{o}^{(i)} \,.
\label{TDhtilde}
\end{eqnarray}
\end{subequations}
As a consequence of equation (\ref{sub}) the new initial conditions on the entrance surface $z=0$ are
\begin{subequations}
\begin{eqnarray}
\tilde{D}_\mathrm{o}^{(i)}(x,0)&=&\exp[-\mathrm{i}f^{(i)}(x,0) 2 K\sin\theta_\mathrm{B}]\Phi(x)\label{eq:o}\\
\tilde{D}_\mathrm{h}^{(i)}(x,0)&=&0\ .
\label{eq:h}
\end{eqnarray}
\end{subequations}
By the Riemann-Green method (Authier \& Simon, 1968,
Takagi, 1969, Sommerfeld, 1964, Palmisano \& Zosi, 2005) we can find a solution by quadrature
for the system (\ref{TDotilde}, \ref{TDhtilde}) with initial conditions (\ref{eq:o}, \ref{eq:h})
\begin{equation}
\tilde{D}_\mathrm{o,h}^{(i)}(x,z)
=
\int_{-\infty}^{+\infty}G_\mathrm{o,h}(x-x',z)
\exp[-\mathrm{i} f^{(i)}(x',0) 2 K \sin\theta_\mathrm{B}]\Phi(x')\,\mathrm{d}x,'
\end{equation}
where the kernels $G_\mathrm{o}$ and $G_\mathrm{h}$ are
\begin{eqnarray}
G_\mathrm{o}(x,z)
&=&
\delta(x+z\tan\theta_\mathrm{B})\nonumber\\
&-&
\frac{K}{4|\sin\theta_\mathrm{B}|}\sqrt{\chi_{\mathrm{h}}\chi_{-\mathrm{h}}\,}\,
\mathrm{H}(z|\tan\theta_\mathrm{B}|+x)\mathrm{H}(z|\tan\theta_\mathrm{B}|-x)\nonumber\\
&\times&
\sqrt{\frac{z\tan\theta_\mathrm{B}-x}{z\tan\theta_\mathrm{B}+x}\,}
\mathrm{J}_1\!\left(
\frac{\mathrm{K}}{2|\sin\theta_\mathrm{B}|}\sqrt{\chi_{\mathrm{h}}\chi_{-\mathrm{h}}\,}\,\sqrt{z^2 \tan^2\theta_\mathrm{B}-x^2\,}\,
\right)
\label{ker-opdo}
\end{eqnarray}
and
\begin{eqnarray}
G_{\mathrm{h}}(x,z)
&=&
\frac{\mathrm{i}}{4}
\frac{K\chi_{\mathrm{h}}}{|\sin\theta_\mathrm{B}|}
\mathrm{H}(z|\tan\theta_\mathrm{B}|+x)\mathrm{H}(z|\tan\theta_\mathrm{B}|-x)
\nonumber\\
&\times&
\mathrm{J}_0\left(\frac{K}{2|\sin\theta_\mathrm{B}|}\sqrt{\chi_{\mathrm{h}}\chi_{-\mathrm{h}}\,}\,
\sqrt{z^2\tan^2\theta_\mathrm{B}-x^2\,}\,\right).
\label{ker-opdh}
\end{eqnarray}
In equations (\ref{ker-opdo}) and (\ref{ker-opdh})
$\mathrm{H}(z)$ is the Heaviside function and $\mathrm{J}_0(z)$ and $\mathrm{J}_1(z)$
are the Bessel functions of the first kind and order $0$, $1$,
respectively.
Therefore the solution to Takagi-Taupin equations (\ref{TDocar}, \ref{TDhcar})
with initial conditions (\ref{ico}, \ref{ich}) is
\begin{eqnarray}
d_\mathrm{o,h}^{(i)}(x,z)
&=&
\exp(\mathrm{i} f^{(i)}(x,z) 2 K \sin\theta_\mathrm{B})\nonumber\\
&\times&
\int_{-\infty}^{+\infty}
G_\mathrm{o,h}(x-x',z)\exp[-\mathrm{i} f^{(i)}(x',0) 2 K \sin\theta_\mathrm{B}]\Phi(x')\,\mathrm{d}x'\,.
\label{solfin}
\end{eqnarray}
We see from equation (\ref{solfin}) that the effect of a constant strain gradient has been reduced to
a similarity transformation of
the $G_\mathrm{o}$ and $G_\mathrm{h}$ kernels (Mana \& Palmisano 2004).
Equations (\ref{subr}) and (\ref{solfin})
show that the intensities of the transmitted and diffracted beams
$D_\mathrm{o}^{(i)}(x,z)$ and $D_\mathrm{h}^{(i)}(x,z)$
on the exit surface $z=T$ are
\begin{eqnarray}
I_\mathrm{o,h}^{(i)}(\alpha)
&=&
\int_{-\infty}^{+\infty}
\left|
D_\mathrm{o,h}^{(i)}(x, T)
\right|^2 \cos\theta_\mathrm{B}\,\mathrm{d}x\nonumber\\
&=&
\int_{-\infty}^{+\infty}
\exp[-K \Im(\chi_\mathrm{o})T/\cos\theta_\mathrm{B}]\nonumber\\
&\times&
\left|
\int_{-\infty}^{+\infty}
G_\mathrm{o,h}(x-x', T)\exp[- \mathrm{i} f^{(i)}(x',0) 2 K \sin\theta_\mathrm{B}]\Phi(x')\,\mathrm{d}x'
\right|^2
\cos\theta_\mathrm{B}\,\mathrm{d}x\,,\nonumber\\
&&
\label{inten}
\end{eqnarray}
where $\Im(\chi_\mathrm{o})$ is the imaginary part of $\chi_\mathrm{o}$.
Eventually, substitution of equation (\ref{fi}) into equation (\ref{inten}) gives
\begin{eqnarray}
&&I_\mathrm{o,h}^{(i)}(\alpha)=
\int_{-\infty}^{+\infty}
\exp[-K \Im(\chi_\mathrm{o})T/\cos\theta_\mathrm{B}]
\left|
\int_{-\infty}^{+\infty}
G_\mathrm{o,h}(x-x', T)
\right.
\nonumber\\
&&\times
\left.
\exp
\left\{
-\mathrm{i} 2 K \sin\theta_\mathrm{B}\frac{\cos\theta_\mathrm{B}}{R_0+z_m}
\left[
\frac{1}{4}g(i)\frac{x'^2}{\sin\theta_\mathrm{B}}
-
\frac{1}{2}x'
\left(
g(i)\frac{z_m}{\cos\theta_\mathrm{B}} - \frac{R_0+z_m}{\sin\theta_\mathrm{B}}\alpha
\right)
\right]
\right\}\,\Phi(x')\mathrm{d}x'
\right|^2
\nonumber\\
&&\times
\cos\theta_\mathrm{B}\,\mathrm{d}x\,.
\label{intenr}
\end{eqnarray}
Equation (\ref{intenr}) gives the rocking curves $I_\mathrm{o,h}^{(i)}(\alpha)$
when the crystal is distorted by the displacement field (\ref{ucard}) or (\ref{ucaru}),
the external crystal surfaces are flat
and $\Phi(x)$ is the complex field amplitude of a generic incoming beam.
Additionally, and generally, the effect of the displacement fields
(\ref{ucard}, \ref{ucaru}) or (\ref{contr}) on the
intensity $I_\mathrm{o,h}^{(i)}$ in equation (\ref{intenr}) is seen to
consist of a phase-redefinition of the initial condition $\Phi(x)$.
Evaluation of equation (\ref{intenr}) in the limit with $R_0$ tending to $+\infty$
gives the rocking curve of a perfect analyzer crystal.
With the variable change $\alpha=\bar{\alpha}^{(i)}+\alpha'$\,, where $\bar{\alpha}^{(i)}$ is
\begin{subequations}
\begin{eqnarray}
\bar{\alpha}^{(i)}
&=&g(i)\frac{z_\mathrm{m}}{R_0+z_\mathrm{m}}\tan\theta_\mathrm{B}\label{bra}\\
&=&
-\frac{\partial u_x^{(i)}}{\partial x}(x,0)\tan\theta_\mathrm{B}\,,
\label{braa}
\end{eqnarray}
\end{subequations}
equation (\ref{intenr}) can be reduced to the simpler form
\begin{eqnarray}
&&I_\mathrm{o,h}^{(i)}(\bar{\alpha}^{(i)}+\alpha')=
\int_{-\infty}^{+\infty}
\exp[-K \Im(\chi_\mathrm{o})T/\cos\theta_\mathrm{B}]
\left|
\int_{-\infty}^{+\infty}
G_\mathrm{o,h}(x-x', T)
\right.
\nonumber\\
&&\times
\left.
\exp
\left\{
-\mathrm{i} 2 K \sin\theta_\mathrm{B}\frac{\cos\theta_\mathrm{B}}{R_0+z_m}
\left[
\frac{1}{4}g(i)\frac{x'^2}{\sin\theta_\mathrm{B}}
+
\frac{1}{2}x'
\frac{R_0+z_m}{\sin\theta_\mathrm{B}}\alpha'
\right]
\right\}\,\Phi(x')\mathrm{d}x'
\right|^2
\cos\theta_\mathrm{B}\,\mathrm{d}x\,.\nonumber\\
&&
\label{intenal}
\end{eqnarray}
%
\subsection{Cylindrical surfaces}\label{curved}
By application of the Riemann-Green method, the solutions of system (\ref{TDocar}, \ref{TDhcar}),
with the initial conditions (\ref{boundDo}) and (\ref{boundDh}),
are the flux integral
\begin{eqnarray}
d_\mathrm{o,h}^{(i)}(x,z)
&=&
\exp[\mathrm{i} f^{(i)}(x,z) 2 K \sin\theta_\mathrm{B}]\nonumber\\
&\times&
\int_{\Gamma}
G_\mathrm{o,h}(x-\Gamma_x,z-\Gamma_z)\exp[- \mathrm{i} f^{(i)}(\Gamma) 2 K \sin\theta_\mathrm{B}]\nonumber\\
&\times&
\exp
\left[
-\mathrm{i}\frac{K\chi_\mathrm{o}}{2}
\frac{\mathbf{\hat{s}}_\mathrm{o}+\mathbf{\hat{s}}_\mathrm{h}}{1+
\mathbf{\hat{s}}_\mathrm{o}\cdot\mathbf{\hat{s}}_\mathrm{h}}
\cdot(\Gamma_x {\bf \hat{x}},\Gamma_z {\bf \hat{z}})
\right]
\Phi(\Gamma)
\frac{\mathbf{\hat{s}}_\mathrm{o}\cdot\mathbf{\hat{n}}_\Gamma}{\cos\theta_\mathrm{B}}
\,\mathrm{d}\Gamma\,,
\label{solfinc}
\end{eqnarray}
which generalizes equation (\ref{solfin}).
The unit vector $\mathbf{\hat{n}}_\Gamma$ is the inward normal to the
$\Gamma:[\tau_1,\tau_2]\rightarrow\mathbf{R}^2$ surface,
$\Gamma_x$ and $\Gamma_z$ are the surface parametric components,
$\mathrm{d}\Gamma$ is a shorthand form for $\|\mathrm{d}\Gamma/\mathrm{d}\tau\|\,\mathrm{d}\tau$,
and $0<\|\mathrm{d}\Gamma/\mathrm{d}\tau\|<1/\tan\theta_\mathrm{B}\, \forall\tau\in[\tau_1,\tau_2]$.
Let $\mathbf{\hat{n}}_\Sigma$ be the outward normal to the exit surface $\Sigma$\,.
From equations (\ref{subr}), (\ref{solfinc}) and (\ref{fi}) the intensity of the
forward transmitted and diffracted beams are
\begin{eqnarray}
&&I_\mathrm{o,h}^{(i)}(\alpha)
=
\int_{\Sigma}
\left|
D_\mathrm{o,h}^{(i)}(\Sigma)
\right|^2
\mathbf{\hat{s}}_\mathrm{h}\cdot\mathbf{\hat{n}}_\Sigma\mathrm{d}\Sigma\nonumber\\
&&
=
\int_{\Sigma}
\exp
\left(
-K\Im(\chi_\mathrm{o})
\frac{\Sigma_z}{\cos\theta_\mathrm{B}}
\right)
\left.
\left|
\int_{\Gamma}
G_\mathrm{o,h}(\Sigma_{x}-\Gamma_{x},\Sigma_{z}-\Gamma_{z})
\right.
\exp
\right\{
-\mathrm{i} 2 K \sin\theta_\mathrm{B}\nonumber\\
&&
\left.
\times
\frac{\sin\theta_\mathrm{B}\cos\theta_\mathrm{B}}{R_0+z_m}
\left[
g(i)\frac{1}{4}
\left(
\frac{\Gamma_{x}}{\sin\theta_\mathrm{B}}+\frac{\Gamma_{z}}{\cos\theta_\mathrm{B}}
\right)^2
-
\frac{1}{2}
\left(
\frac{\Gamma_{x}}{\sin\theta_\mathrm{B}}+\frac{\Gamma_{z}}{\cos\theta_\mathrm{B}}
\right)
\left(g(i)\frac{z_m}{\cos\theta_\mathrm{B}}-\frac{R_0+z_m}{\sin\theta_\mathrm{B}}\alpha\right)
\right]
\right\}
\nonumber\\
&&
\times
\left.
\exp
\left(
-\mathrm{i}\frac{K\chi_\mathrm{o}}{2}
\frac{\Gamma_z}{\cos\theta_\mathrm{B}}
\right)
\Phi(\Gamma_x,\Gamma_z)
\frac{\mathbf{\hat{s}}_\mathrm{o}\cdot\mathbf{\hat{n}}_\Gamma}{\cos\theta_\mathrm{B}}\,\mathrm{d}\Gamma
\right|^2
\mathbf{\hat{s}}_\mathrm{h}\cdot\mathbf{\hat{n}}_\Sigma\,\mathrm{d}\Sigma\,.
\label{intenc}
\end{eqnarray}

In the concave case, 
the equation (\ref{intenc}) with use of the  
$\bar{\alpha}^{(1)}$, and the definitions
(\ref{bra}) and (\ref{GammaR}-\ref{Sigman}), can be rewritten as
\begin{eqnarray}
&&I_\mathrm{o,h}^{(1)}(\bar{\alpha}^{(1)}+\alpha')
=
\int_{\sigma_1}^{\sigma_2}
\exp
\left[
-K\Im(\chi_\mathrm{o})
\frac{(R_0+T)\cos\sigma-R_0}{\cos\theta_\mathrm{B}}
\right]
\nonumber\\
&&
\times
\left|
\int_{\tau_1}^{\tau_2}
G_\mathrm{o,h}((R_0+T)\sin\sigma-R_0\sin\tau,(R_0+T)\cos\sigma-R_0\cos\tau)
\right.
\nonumber\\
&&
\times
\exp
\left\{
-\mathrm{i} 2 K \sin\theta_\mathrm{B} \frac{\sin\theta_\mathrm{B}\cos\theta_\mathrm{B}}{R_0+z_m}R_0
\left(
\frac{\sin\tau}{\sin\theta_\mathrm{B}}+\frac{\cos\tau-1}{\cos\theta_\mathrm{B}}
\right)
\right.
\nonumber\\
&&
\times
\left.
\left[
g(1)\frac{R_0}{4}
\left(
\frac{\sin\tau}{\sin\theta_\mathrm{B}}+\frac{\cos\tau-1}{\cos\theta_\mathrm{B}}
\right)
+
\frac{1}{2}
\frac{R_0+z_m}{\sin\theta_\mathrm{B}}\alpha'
\right]
\right\}
\nonumber\\
&&
\times
\exp
\left(
-\mathrm{i}\frac{K\chi_\mathrm{o}}{2}
R_0
\frac{\cos\tau-1}{\cos\theta_\mathrm{B}}
\right)
\Phi(R_0\sin\tau,R_0(\cos\tau-1))
\nonumber\\
&&
\left.
\times
\frac{\cos(\theta_\mathrm{B}+\tau)}{\cos\theta_\mathrm{B}}R_0\mathrm{d}\tau
\right|^2
\cos(\theta_\mathrm{B}-\sigma)(R_0+T)\mathrm{d}\sigma\,.
\label{intencr}
\end{eqnarray}

In the convex case,
the equation (\ref{intenc}), by following the same procedure,
that is, with the use of definitions (\ref{bra})
and (\ref{GammaRT}-\ref{SigmanT}), can be rewritten as
\begin{eqnarray}
&&I_\mathrm{o,h}^{(2)}(\bar{\alpha}^{(2)}+\alpha')
=
\int_{\sigma_1}^{\sigma_2}
\exp
\left[
-K\Im(\chi_\mathrm{o})
\frac{R_0+T-R_0\cos\sigma}{\cos\theta_\mathrm{B}}
\right]
\nonumber\\
&&
\times
\left|
\int_{\tau_1}^{\tau_2}
G_\mathrm{o,h}(R_0\sin\sigma-(R_0+T)\sin\tau,R_0\cos\sigma-(R_0+T)\cos\tau)
\right.
\nonumber\\
&&
\times
\exp
\left\{
-\mathrm{i} 2 K \sin^2\theta_\mathrm{B}\cos\theta_\mathrm{B}\frac{R_0+T}{R_0+z_m}
\left(
\frac{\sin\tau}{\sin\theta_\mathrm{B}}+\frac{\cos\tau-1}{\cos\theta_\mathrm{B}}
\right)
\right.
\nonumber\\
&&
\times
\left.
\left[
g(2)\frac{R_0+T}{4}
\left(
\frac{\sin\tau}{\sin\theta_\mathrm{B}}+\frac{\cos\tau-1}{\cos\theta_\mathrm{B}}
\right)
+
\frac{1}{2}
\frac{R_0+z_m}{\sin\theta_\mathrm{B}}\,\alpha'
\right]
\right\}
\nonumber\\
&&
\times
\exp
\left[
-\mathrm{i}\frac{K\chi_\mathrm{o}}{2}
(R_0+T)\frac{\cos\tau-1}{\cos\theta_\mathrm{B}}
\right]
\Phi((R_0+T)\sin\tau,(R_0+T)(1-\cos\tau))
\nonumber\\
&&
\left.
\times
\frac{\cos(\theta_\mathrm{B}-\tau)}{\cos\theta_\mathrm{B}}(R_0+T)\mathrm{d}\tau
\right|^2
\cos(\theta_\mathrm{B}+\sigma)R_0\mathrm{d}\sigma\,.
\label{intencrT}
\end{eqnarray}

\section{Double crystal diffractometer} \label{double}
\begin{figure}
\begin{minipage}[t]{0.45\textwidth}
\begin{center}
\includegraphics[width=6.0 cm]{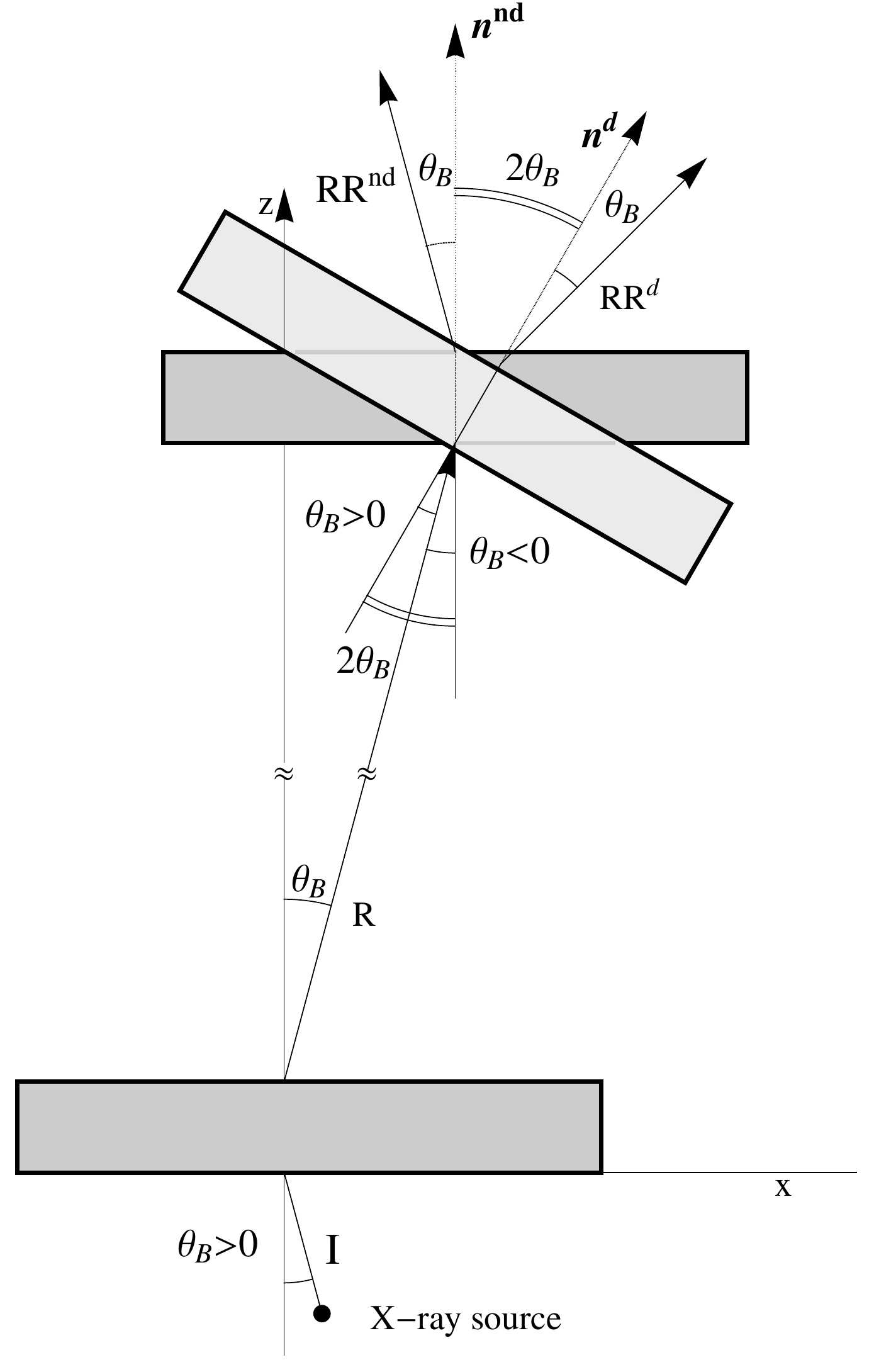}
\end{center}
\end{minipage}
\hfill
\begin{minipage}[t]{0.45\textwidth}
\begin{center}
\includegraphics[width=6.0 cm]{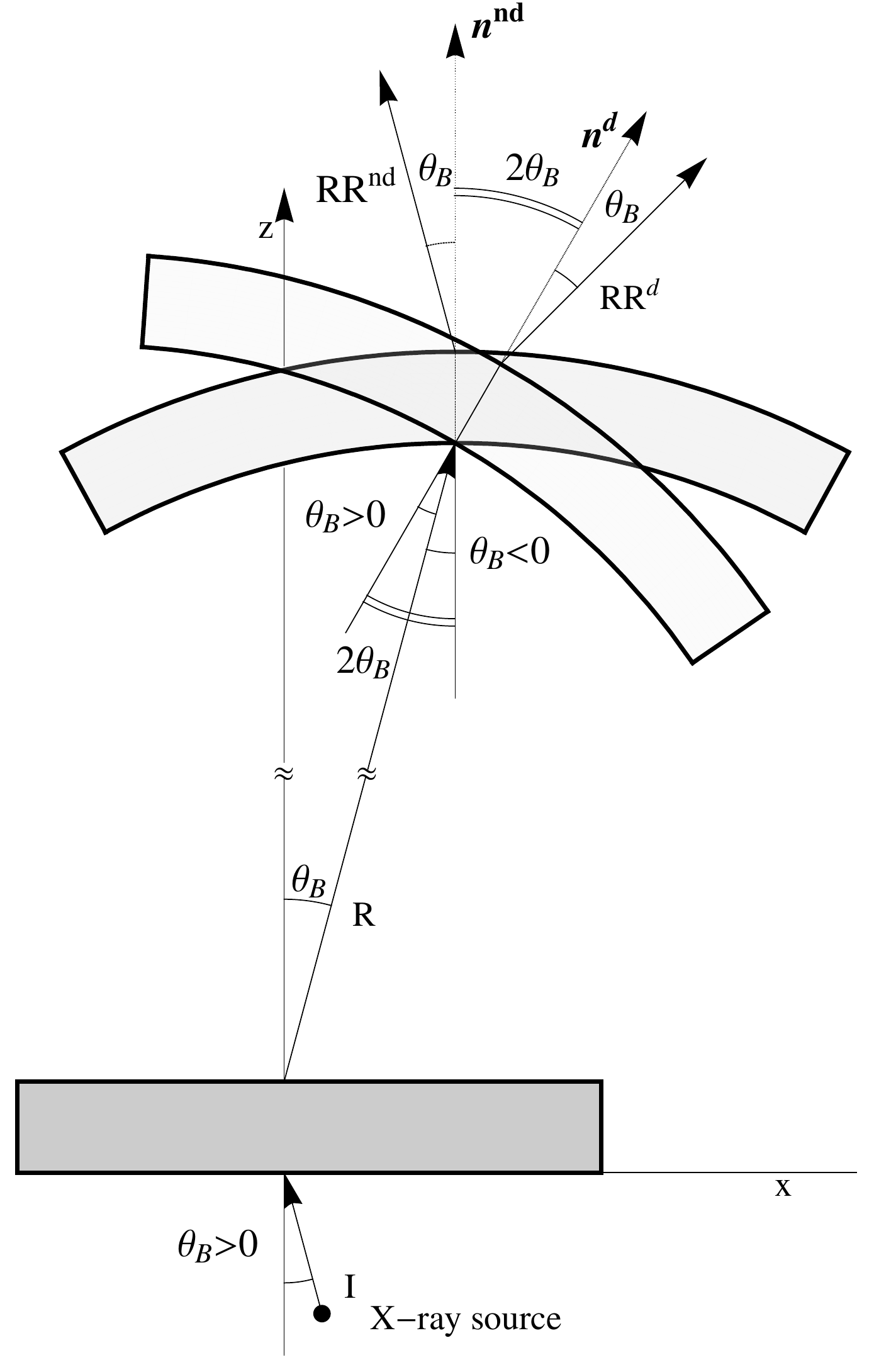}
\end{center}
\end{minipage}
\caption{
Layout of a two-crystal diffractometer in non-dispersive and dispersive geometries. 
The symbol I is the incident beam, R is the beam reflected by the collimating crystal,
 RR$^{nd}$ and RR$^d$ are the double reflected outgoing beams,
the superscripts ``nd'' and ``d'' refer to the non-dispersive
and dispersive setup. The unit vectors ${\bf \hat{s}_{o,h}}$ are defined 
in equations (\ref{veccar}) and (\ref{veccarh}).
The angle of incidence $\theta$  on the collimating crystal
is positive ($\theta=\theta_\mathrm{B}$);
the angle of incidence $\theta$  on the analyzer crystal is negative
for the non-dispersive setup and positive for the dispersive setup.
}
\label{f2}

\end{figure}

Figs.~\ref{f1} and \ref{f2} show the Laue-Laue diffractometer in
both the non-dispersive and dispersive setups.
We consider a flat collimating perfect-crystal and a monochromatic point
source located in $(L\sin\theta_\mathrm{B}, -L\cos\theta_\mathrm{B})$,
where $L$ is the distance between the source and the entrance point of the collimating crystal.
We have chosen the vector $\mathbf{h}_0$ of the analyzer reference
perfect-lattice equal to the vector $\mathbf{h}_0$ collimating crystal;
therefore, these two lattices have the same spacing and,
when $\alpha=0$, they are parallel. If the two crystals have the same lattice spacing,
the analyzer rotation $\alpha$ between the non-dispersive and dispersive geometries is equal
to $2 \theta_\mathrm{B}$; a different rotation is related to a different lattice
spacing in the collimating and analyzer crystals.
\subsection{Rocking curves} \label{RC}
Let us confine our study to the reflected beam. When we examine the equation (\ref{intenal}),
we see that, if
\begin{equation}
\Phi(x)=\Phi(-x)\,,
\label{sym}
\end{equation}
where $\Phi(x)$ is the amplitude of the o-component of the external field on
the entrance surface of the analyzer,
the intensity profile of
$I_\mathrm{h}^{(i)}(\bar{\alpha}^{(i)}+\alpha')$ is invariant under the
$\alpha'\rightleftarrows-\alpha'$ exchange, as
can be easily checked by the double substitution $x'=-\tilde x'\,, x=-\tilde x$\,.
This means that $I_\mathrm{h}^{(i)}(\alpha)$
has a vertical symmetry-axis passing through $\bar{\alpha}^{(i)}$.
Let us note that equation (\ref{sym}) is fulfilled if the amplitude of the o-component of
the incoming beam is an even function.

The formula $\bar{\alpha}^{(i)}/\tan\theta_\mathrm{B}=-\partial u_x^{(i)}/\partial x$,
obtained from equation (\ref{braa}), is
our formulation of the $\Delta\theta/\tan\theta_\mathrm{B}=-\Delta d/d$ Bragg's law.
It is to be noted that, if the entrance-surface displacement field is zero, i.e.,
when $z_\mathrm{m}=0$, then $\bar{\alpha}^{(i)}=0$ and there is no shift
of the reflection peak with respect to the perfect-crystal case.

With crystals having cylindrical surfaces, as long as the approximations
$\cos\tau$
$\approx \cos(\theta_\mathrm{B}+\tau)$
$\approx \cos(\theta_\mathrm{B}-\tau)$
$\approx \cos(\theta_\mathrm{B}-\sigma)$
$\approx \cos(\theta_\mathrm{B}+\sigma)$
$\approx 1$
in equation (\ref{intencr}) and in equation (\ref{intencrT}) are valid,
and the symmetry requirement in equation (\ref{sym}) is satisfied,
the same conclusion holds. In fact $I_\mathrm{h}^{(i)}(\alpha)$, now given by
equation (\ref{intencr}) or equation (\ref{intencrT}) and depending on the concave or convex case,
has a symmetry axis passing through
$\bar{\alpha}^{(i)}=g(i)\tan\theta_\mathrm{B}\,z_\mathrm{m}/(R_0+z_\mathrm{m})$.
The approximations mean that we have small Bragg angles, 
great curvature radii and
the profile of the wave from the collimating crystal is not altered by the geometry of the analyzer crystal.
In Mana {\it et al.} (2004), it is reported that, in the presence of a constant strain
gradient in the analyzer, the Laue-Laue rocking curve is shifted by $(\Delta d/d)\tan\theta_B$,
where the lattice strain is evaluated on the crystal surface. However, this peak shift is not easily measurable.
Since the analyzer rotation between dispersive and non-dispersive
reflection peaks is an experimentally observable quantity, we give now the relevant equation.
In our formalism, the o-component of the analyzer crystal field co-propagates or
counter-propagates with respect to $x$ axis according to the $\theta_\mathrm{B}$ sign;
therefore, the exchange between the dispersive and non-dispersive geometries corresponds
to the substitution of $-\theta_\mathrm{B}$ for $\theta_\mathrm{B}$.
If we observe that the non-dispersive rocking curve peaks when $\alpha=\bar{\alpha}^{(i)}(-\theta_\mathrm{B})$,
where $\theta_\mathrm{B}$ is the Bragg angle for the collimating crystal (Fig.~\ref{f1}),
and we observe as well that the dispersive one peaks when
$\alpha = 2 \theta_\mathrm{B}+\bar{\alpha}^{(i)}(\theta_\mathrm{B})$
(Fig.~\ref{f2}), the sought formula is
\begin{subequations}
\begin{eqnarray}
 \Delta\alpha^{(i)}
 &=&
 2\theta_\mathrm{B}+\bar{\alpha}^{(i)}(\theta_\mathrm{B})-\bar{\alpha}^{(i)}(-\theta_\mathrm{B})
 \label{bra1}\\
 &=&
 2\left[\theta_\mathrm{B}+\bar{\alpha}^{(i)}(\theta_\mathrm{B})\right]
 \label{bra2}\\
 &=&
 2\left[ \theta_\mathrm{B}-\frac{\partial u_x^{(i)}}{\partial x}(x,0)\tan\theta_\mathrm{B} \right]\,,
 \label{bra3}
\end{eqnarray}
\end{subequations} 
where $(\partial u_x^{(i)}/\partial x)(x,0)$ is the lattice strain on the
analyzer entrance surface.
\section{Numerical simulation}\label{num-sim}
In order to validate our previous results we have also studied the diffractometer
operation by solving the Takagi-Taupin equations numerically.
Again we consider the two distinct cases of flat or cylindrical analyzer surfaces.
The collimating crystal is a parallel-sided
silicon slab limited by two surfaces orthogonal to the $(220)$ Bragg planes.
The x- or $\gamma$-ray source illuminates the collimating crystal by
a monochromatic cylindrical wave $D_e(x,z)=\Psi(x,z)\exp(i\mathbf{K}_\mathrm{o}\cdot \mathbf{r})$, where
\begin{equation}
\Psi(x,0)
=
\left\{
\begin{array}{ll}
\displaystyle{\Lambda(x)\frac{[x-(w/2)]^{8} [x+(w/2)]^{8}}{(w/2)^{16}}} &\mbox{if} -w/2 \leqslant x \leqslant w/2 \\
0                                                                 &\mbox{otherwise}
\end{array} \right. ,
\label{fie}
\end{equation}
\begin{equation}
\Lambda(x)=\frac{1}{4 \pi L}
\exp\left(\mathrm{i}\frac{2\pi}{\lambda}\frac{\cos^2{\theta_\mathrm{B}}}{2 L}x^2\right),
\end{equation}
$\lambda=hc/E$ (with $E$ the photon energy) is the wavelength,
$w/2$ is the half-width on the entrance slit, and the Bragg angle is positive.
In the numerical simulation, we considered the two sets of parameters shown in Table 1;
the lower energy value refers to the experimental setup described in Mana \emph{et al.} (2004),
the upper in Massa \emph{et al.}~(2006).
We have considered silicon (220) Bragg planes and  have taken the values of the
dielectric susceptibilities from the Sergey
Stepanov's X-Ray Server http://sergey.gmca.aps.anl.gov~.

As a first step we solved equations (\ref{TDocar}, \ref{TDhcar}) for a perfect crystal
with boundary conditions (\ref{fie}). Subsequently, the reflected beam,
which we shall indicate by $D_\mathrm{h}^{(i),\,\mathrm{col}}(x,T)$, was free-propagated rigidly
from the collimating crystal onto the entrance surface of the analyzer crystal.
\subsection{Flat analyzer surfaces}  \label{flat-anal}
\begin{figure}
\begin{minipage}[t]{0.45\textwidth}
\begin{center}
\includegraphics[width=7.0 cm]{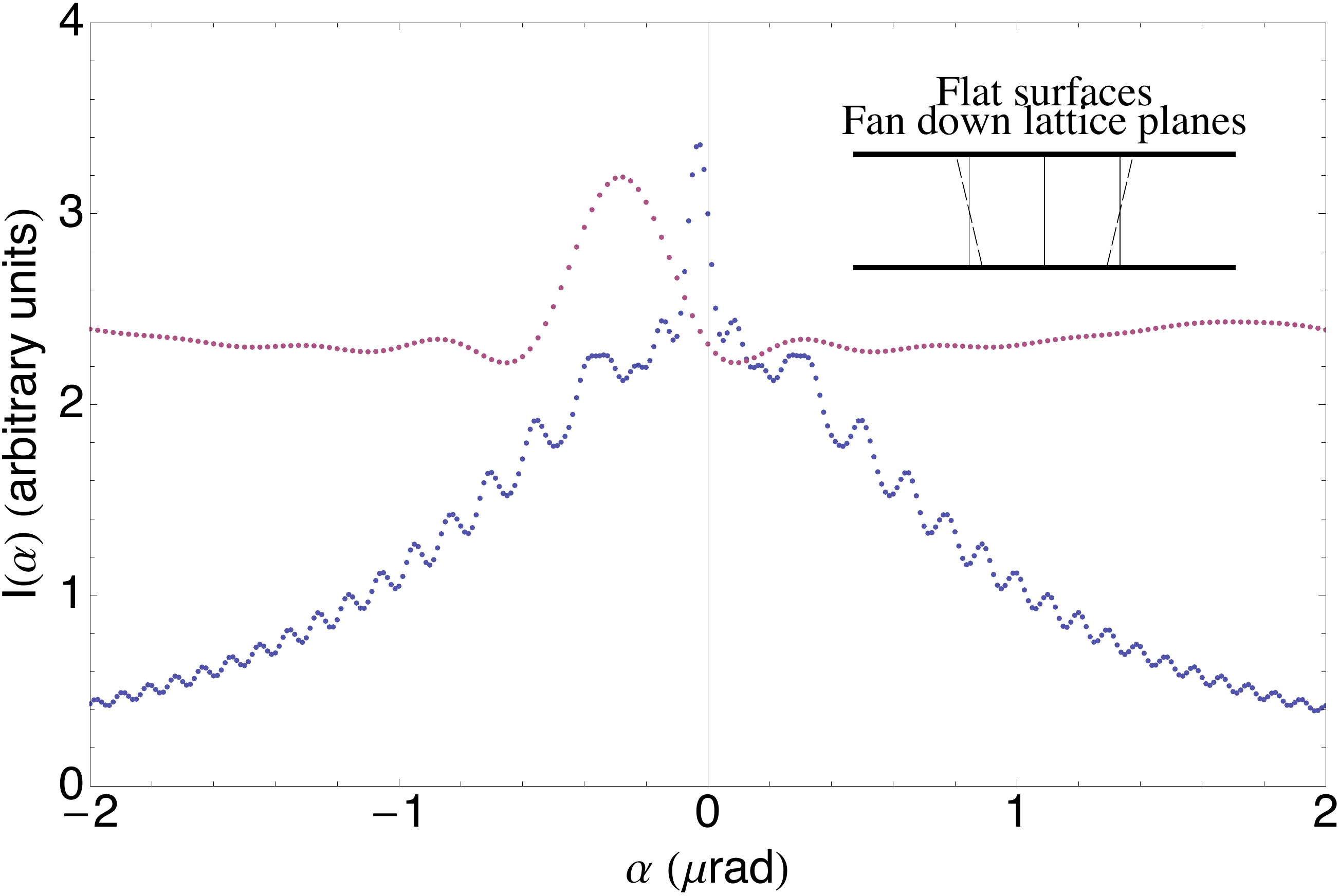}
\end{center}
\end{minipage}
\hfill
\begin{minipage}[t]{0.45\textwidth}
\begin{center}
\includegraphics[width=7.0 cm]{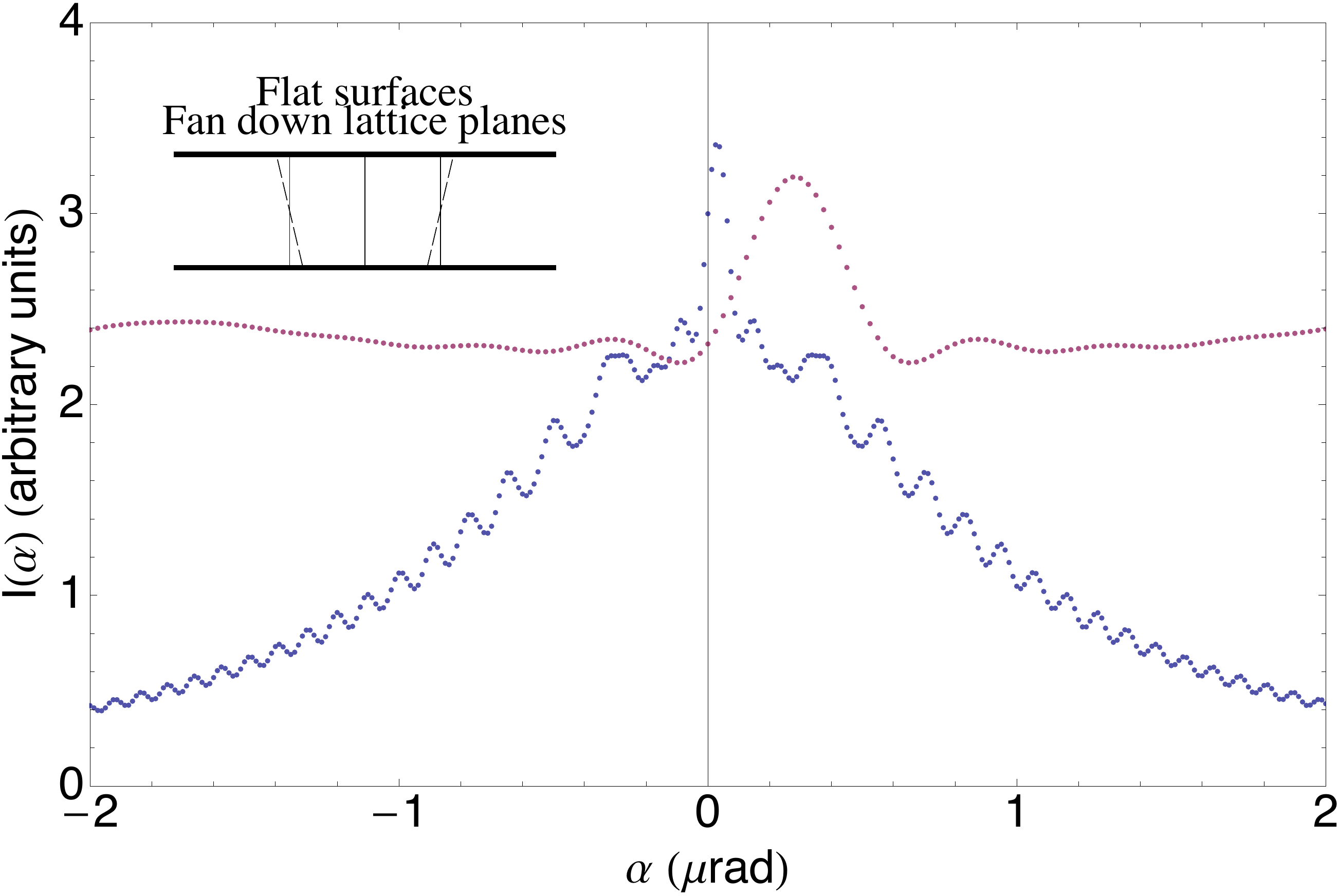}
\end{center}
\end{minipage}
\caption{
Non-dispersive (left) and dispersive (right) 
rocking curves for a flat Si analyzer with a fan-down distortion.
Upper and lower curves refer to the energies of $17$ keV and $184$ keV, respectively.
Bragg planes are (220), the relevant parameter values are given in Table 1.
}
\label{pianaconcavd}
\end{figure}

\begin{figure}
\begin{minipage}[t]{0.45\textwidth}
\begin{center}
\includegraphics[width=7. cm]{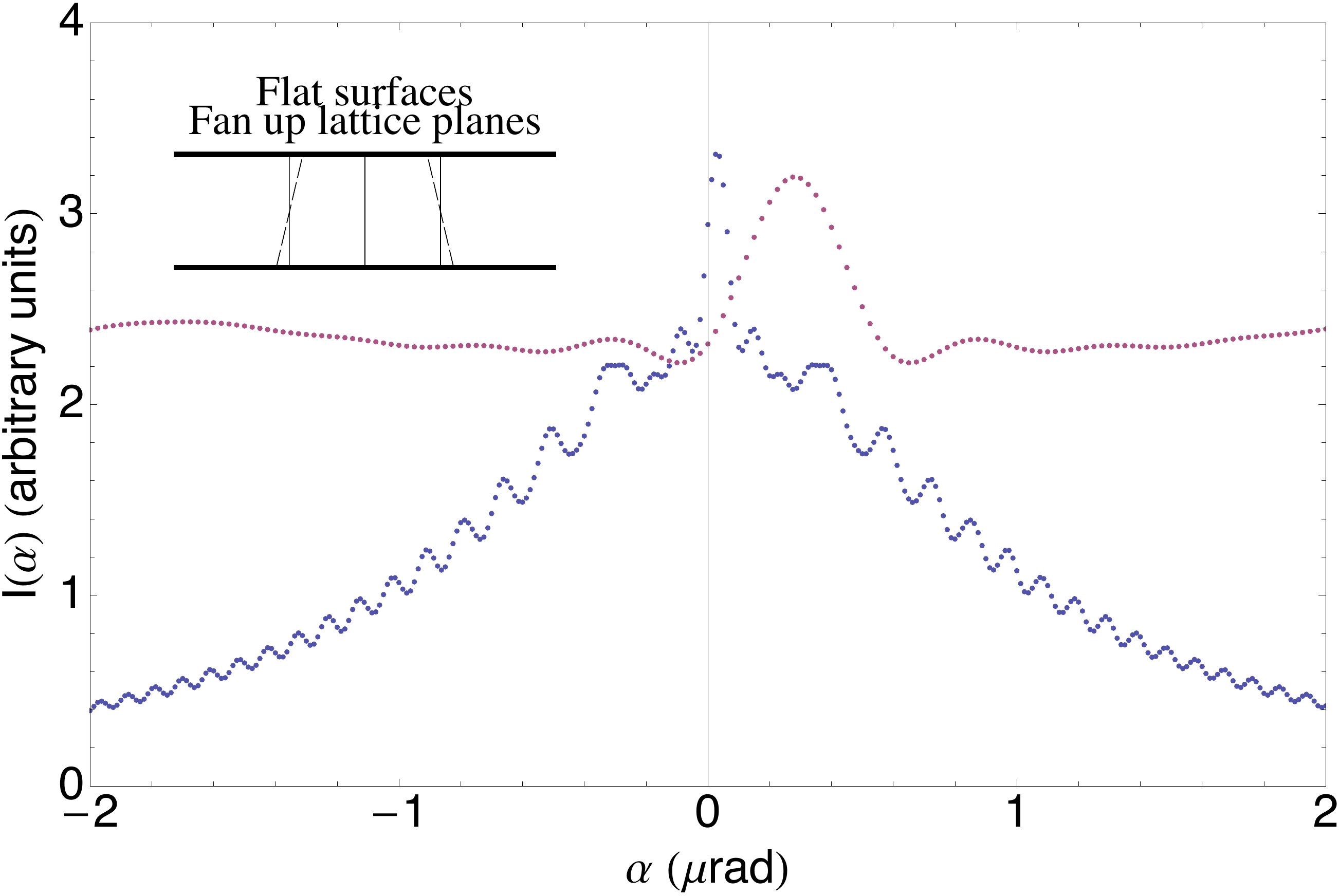}
\end{center}
\end{minipage}
\hfill
\begin{minipage}[t]{0.45\textwidth}
\begin{center}
\includegraphics[width=7. cm]{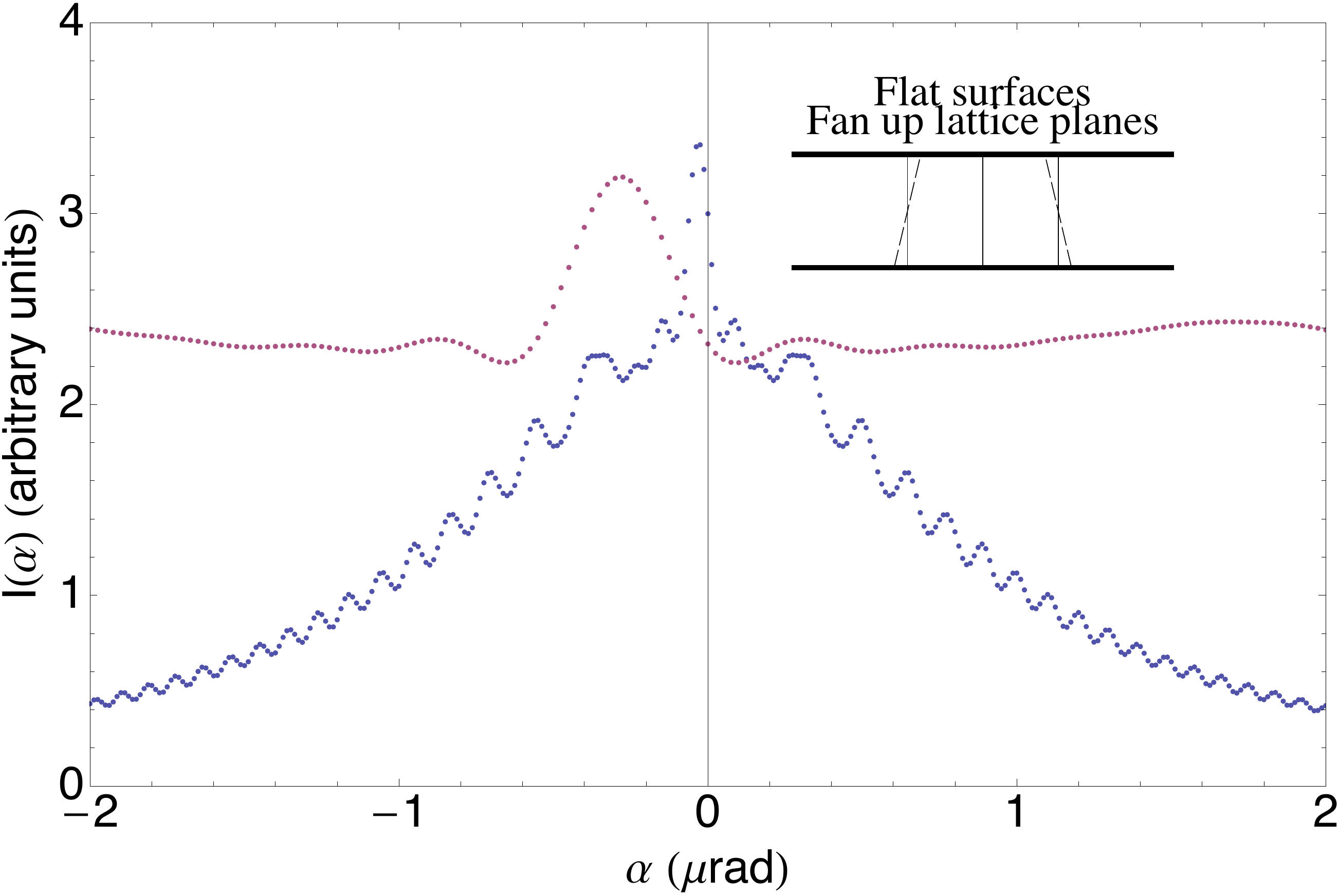}
\end{center}
\end{minipage}
\caption{
Non-dispersive (left) and dispersive (right)
rocking curves for a flat Si analyzer with a fan-up distortion.
Upper and lower curves refer to the energies of $17$ keV and $184$ keV, respectively.
Bragg planes are (220), the relevant parameter values are given in Table 1.
}
\label{pianaconvexd}
\end{figure}

In this case, the propagation equations of the analyzer
fields $D_\mathrm{o}^{(i),\,\mathrm{ana}}(x,z)$ and $D_\mathrm{h}^{(i),\,\mathrm{ana}}(x,z)$
are (\ref{TDocar}, \ref{TDhcar}), where $u_{x}^{(i)}(x,z)$ is
the displacement field in equations (\ref{ucard}) and (\ref{ucaru}), the Bragg angle is negative, and
the initial field values are
\begin{subequations}
\begin{eqnarray}
D_\mathrm{o}^{(i),\,\mathrm{ana}}(x,0)&=&D_\mathrm{h}^{(i),\,\mathrm{col}}(x,T)\label{icos}\\
D_\mathrm{h}^{(i),\,\mathrm{ana}}(x,0)&=&0\label{ichs}\ .
\end{eqnarray}
\end{subequations}
We have calculated the numerical solutions with the aid of
\textsl{MATHEMATICA} (Versions 5.2 and 6.1, Wolfram Res. Inc.);
we have obtained the same results (to within $0.1\%$) by performing the integration
in equation (\ref{solfin}) numerically.
In Figs.~\ref{pianaconcavd} and \ref{pianaconvexd} we show the rocking
curves when the analyzer distortion corresponds to the fan-down and fan-up cases,
in both the non-dispersive and dispersive geometries;
the peak-shifts agree with the values predicted by equation (\ref{bra}).

\begin{figure}
\begin{center}
\includegraphics[width=12.0 cm]{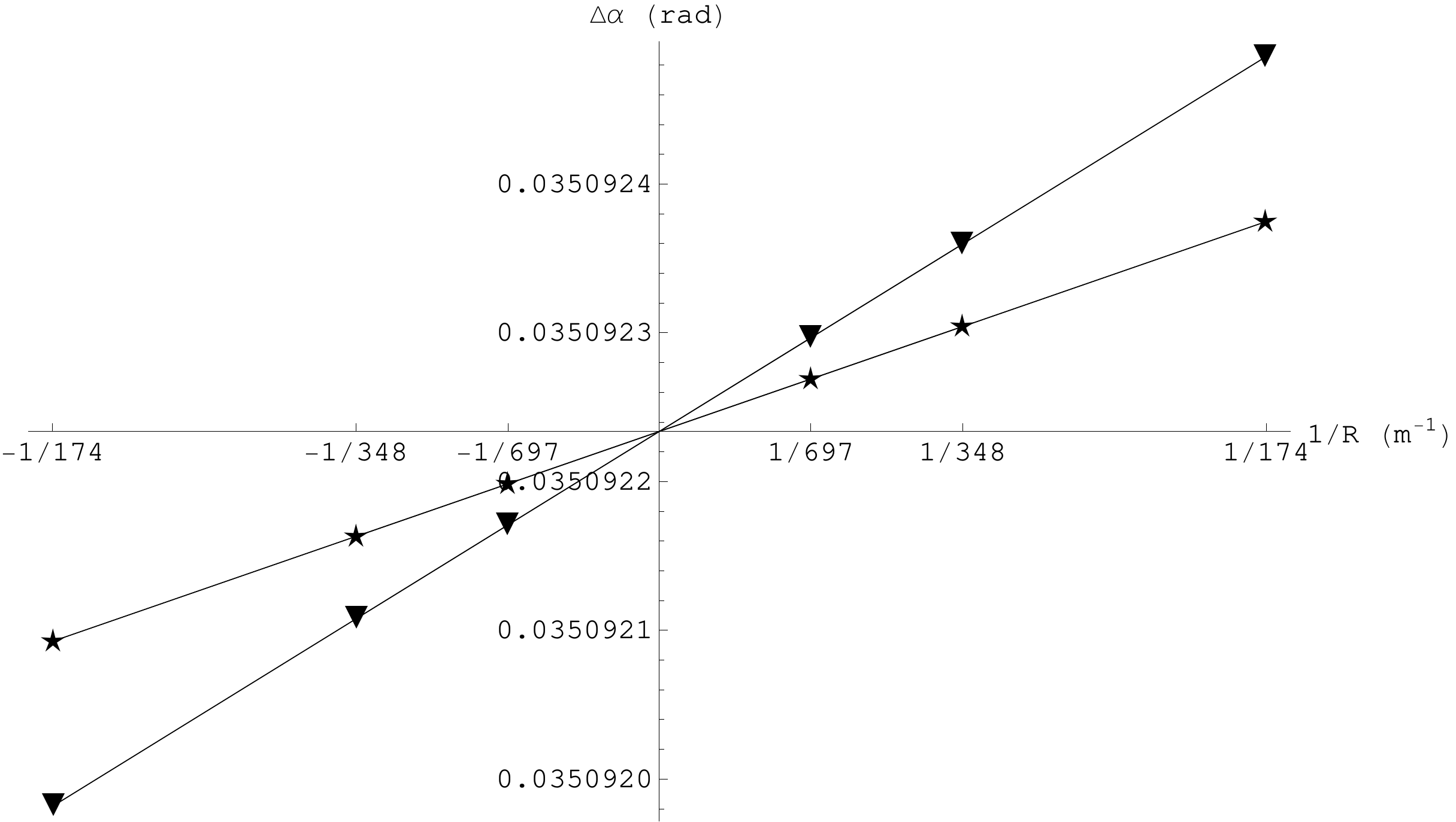}
\end{center}
\caption{
Analyzer rotation $\Delta\alpha$ from non-dispersive to dispersive geometry 
as a function of curvature $1/R$ for $T=1.4$ mm ($\blacktriangledown$), $T=2.5$ mm ($\star$) when $E=184$ keV and $z_\mathrm{m}=T/2$. 
Positive and negative values of $1/R$ refer to the concave and convex cases, respectively, and have been calculated numerically.
}
\label{thetaconf}
\end{figure}

Two cases, calculated numerically
and according to equation (\ref{bra}), are compared in Fig.~\ref{thetaconf},
showing that the maximum difference in $\Delta\alpha$ amounts to a few parts per $10^{-7}$.
Figs.~\ref{pianaconcavd} and \ref{pianaconvexd} require
a few comments.
Firstly, in contrast to experimental observations, there is not spreading in Fig.~\ref{pianaconvexd};
the reason is that we assumed the source monochromatic and we did not integrate over its linewidth.

In a non-dispersive geometry, with a perfect-crystal analyzer having the same lattice spacing as the collimating crystal,
all rocking curves peak when $\alpha=0$, no matter which the
wavelength might be.
On the contrary, in a dispersive geometry, the rocking curves peak when $\alpha=2\theta_\mathrm{B}(\lambda)$ thus
giving rise to a convolution integral.
Secondly, we did not give particular attention to the crystal field intensities;
therefore, the relative intensities of the $17$ keV and $184$ keV plots are meaningless and the two curves are not comparable.
Finally, the dotted lines show only the central part of the extremely wide $17$ keV curve.
\subsection{Cylindrical analyzer surfaces}
In this case, we rewrote the Takagi-Taupin equations (\ref{TDocar}, \ref{TDhcar})
in polar coordinates, $\rho = \sqrt{x^2+z^2}$ and
$\varphi=\arctan(z/x)$, the reference-frame origin being in the centre
of curvature of the entrance surface. Hence,
\begin{subequations}
\begin{eqnarray}
&&
\cos{(\theta_\mathrm{B}+\varphi)}\frac{\partial D_\mathrm{o}^{(i),\,\mathrm{ana}} }{\partial \rho}
-
\frac{\sin{(\theta_\mathrm{B}+\varphi)}}{\rho}\frac{\partial D_\mathrm{o}^{(i),\,\mathrm{ana}} }{\partial \varphi}
=
\mathrm{i}\frac{K\chi_{-\mathrm{h}}}{2} D_\mathrm{h}^{(i),\,\mathrm{ana}}
\label{TDopol}\\
&&
\cos{(\theta_\mathrm{B}-\varphi)}\frac{\partial D_\mathrm{h}^{(i),\,\mathrm{ana}} }{\partial \rho}
+
\frac{\sin{(\theta_\mathrm{B}-\varphi)}}{\rho}
\frac{\partial D_\mathrm{h}^{(i),\,\mathrm{ana}} }{\partial \varphi}
=
\mathrm{i}\frac{K\chi_{\mathrm{h}}}{2} D_\mathrm{o}^{(i),\,\mathrm{ana}}
\nonumber \\
&&
\ +
\mathrm{i} 2 K\sin{\theta_\mathrm{B}}
\left(
\cos{(\theta_\mathrm{B}-\varphi)}
\frac{\partial u_x^{(i)} }{\partial \rho}
+
\frac{\sin{(\theta_\mathrm{B}-\varphi)}}{\rho}
\frac{\partial u_x^{(i)} }{\partial \varphi}
\right)
D_\mathrm{h}^{(i),\,\mathrm{ana}}
\label{TDhpol}
\end{eqnarray}
\end{subequations}
the $x$-component of the displacement field,
including the rotation term, is
\begin{equation}
u_{x,d}^{(i)}(\rho,\varphi)
=
\rho\sin{\varphi}\left(\frac{\rho\cos{\varphi}-R_0-z_\mathrm{m}}{R_0+z_\mathrm{m}}\right)
+
g(i)
\alpha\rho\cos{\varphi}
\label{eq:sts}
\end{equation}
and the boundary conditions for the concave case are
\begin{subequations}
\begin{eqnarray}
D_\mathrm{o}^{(i),\,\mathrm{ana}}(R_0,\tau)&=&D_\mathrm{h}^{(i),\,\mathrm{col}}
(R_0\sin\tau+R_0(1-\cos\tau)\tan\theta_\mathrm{B}, T)
\label{icosr}\\
D_\mathrm{h}^{(i),\,\mathrm{ana}}(R_0,\tau)&=&0
\label{ichsr},
\end{eqnarray}
\end{subequations}
and
\begin{subequations}
\begin{eqnarray}
D_\mathrm{o}^{(i),\,\mathrm{ana}}(R_0+T,\tau)&=&D_\mathrm{h}^{(i),\,\mathrm{col}}
((R_0+T)\sin\tau-(R_0+T)(1-\cos\tau)\tan\theta_\mathrm{B}, T)
\label{icosrc}\\
D_\mathrm{h}^{(i),\,\mathrm{ana}}(R_0+T,\tau)&=&0
\label{ichsrc},
\end{eqnarray}
\end{subequations}
for the convex case. Here, too, the numerical solution
of equations (\ref{TDopol}) and (\ref{TDhpol})
agrees with the numerical integration of equation (\ref{solfinc}).

\begin{figure}
\begin{minipage}[t]{0.45\textwidth}
\begin{center}
\includegraphics[width=7.0 cm]{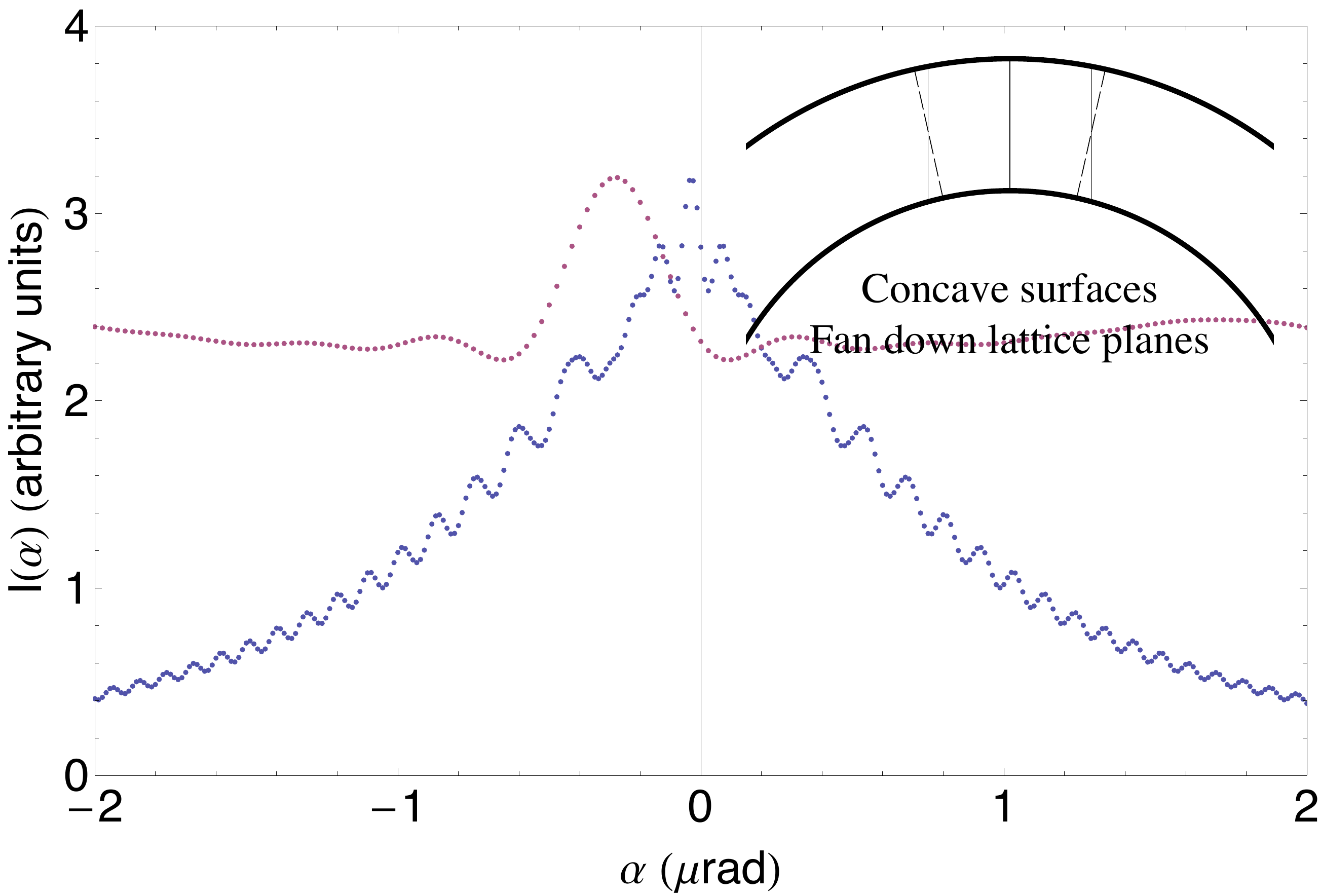}
\end{center}
\end{minipage}
\hfill
\begin{minipage}[t]{0.45\textwidth}
\begin{center}
\includegraphics[width=7.0 cm]{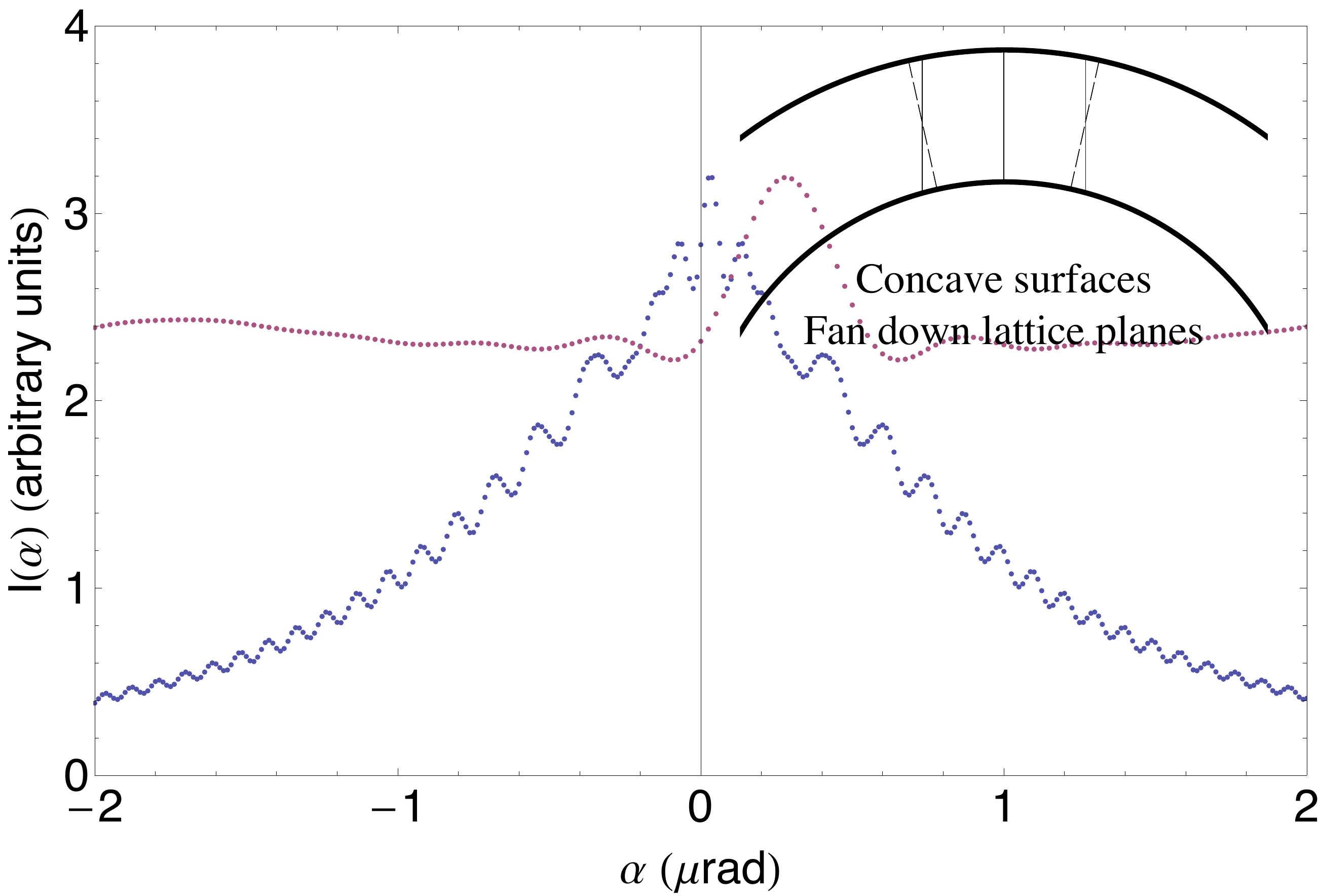}
\end{center}
\end{minipage}
\caption{
Non-dispersive (left) and dispersive (right)
rocking curves for a concave Si analyzer with a fan-down distortion.
Upper and lower curves refer to the energies of $17$ keV and $184$ keV, respectively.
Bragg planes are (220), the relevant parameter values are given in Table 1.
}
\label{concavd}
\end{figure}

\begin{figure}
\begin{minipage}[t]{0.45\textwidth}
\begin{center}
\includegraphics[width=7.0 cm]{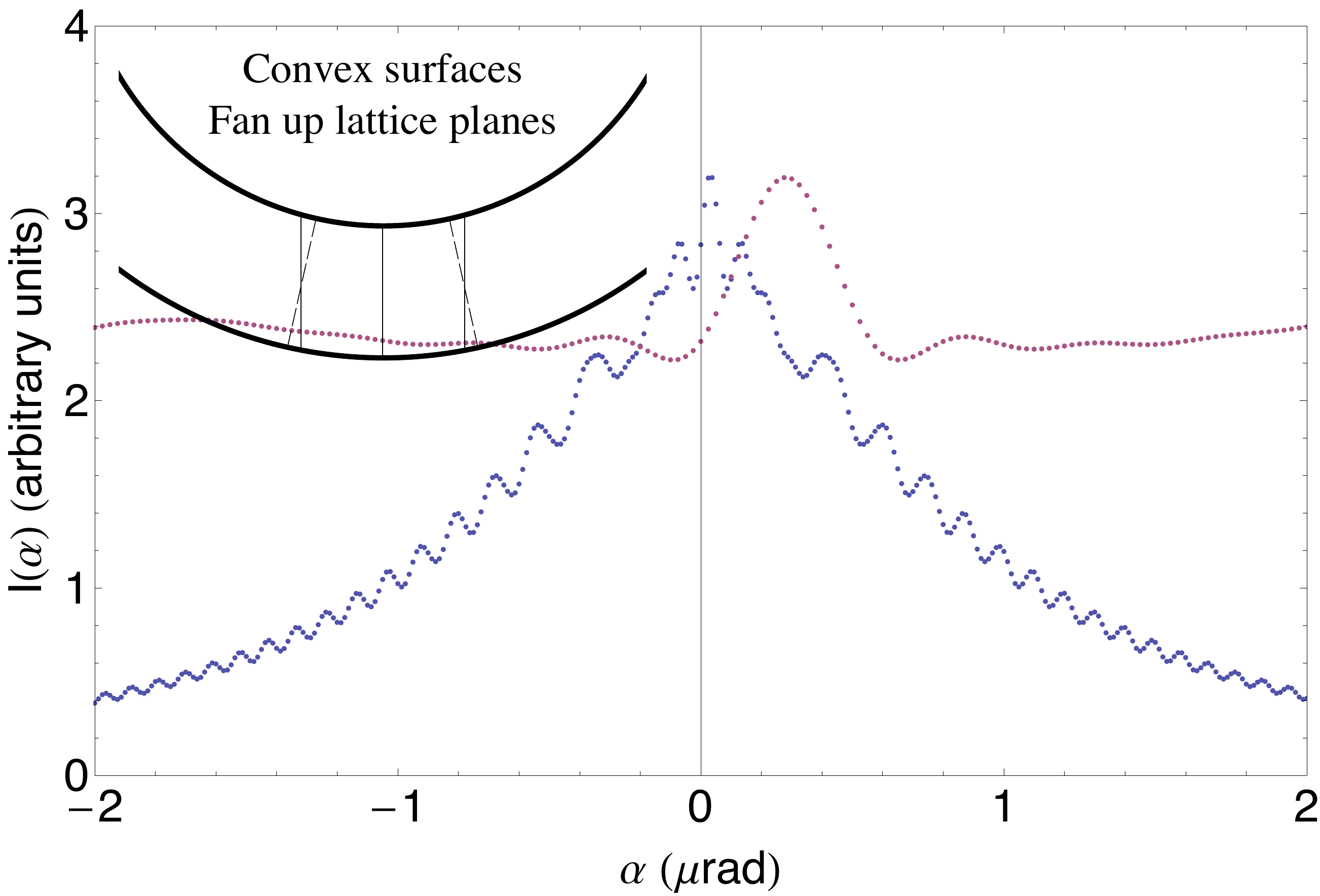}
\end{center}
\end{minipage}
\hfill
\begin{minipage}[t]{0.45\textwidth}
\begin{center}
\includegraphics[width=7.0 cm]{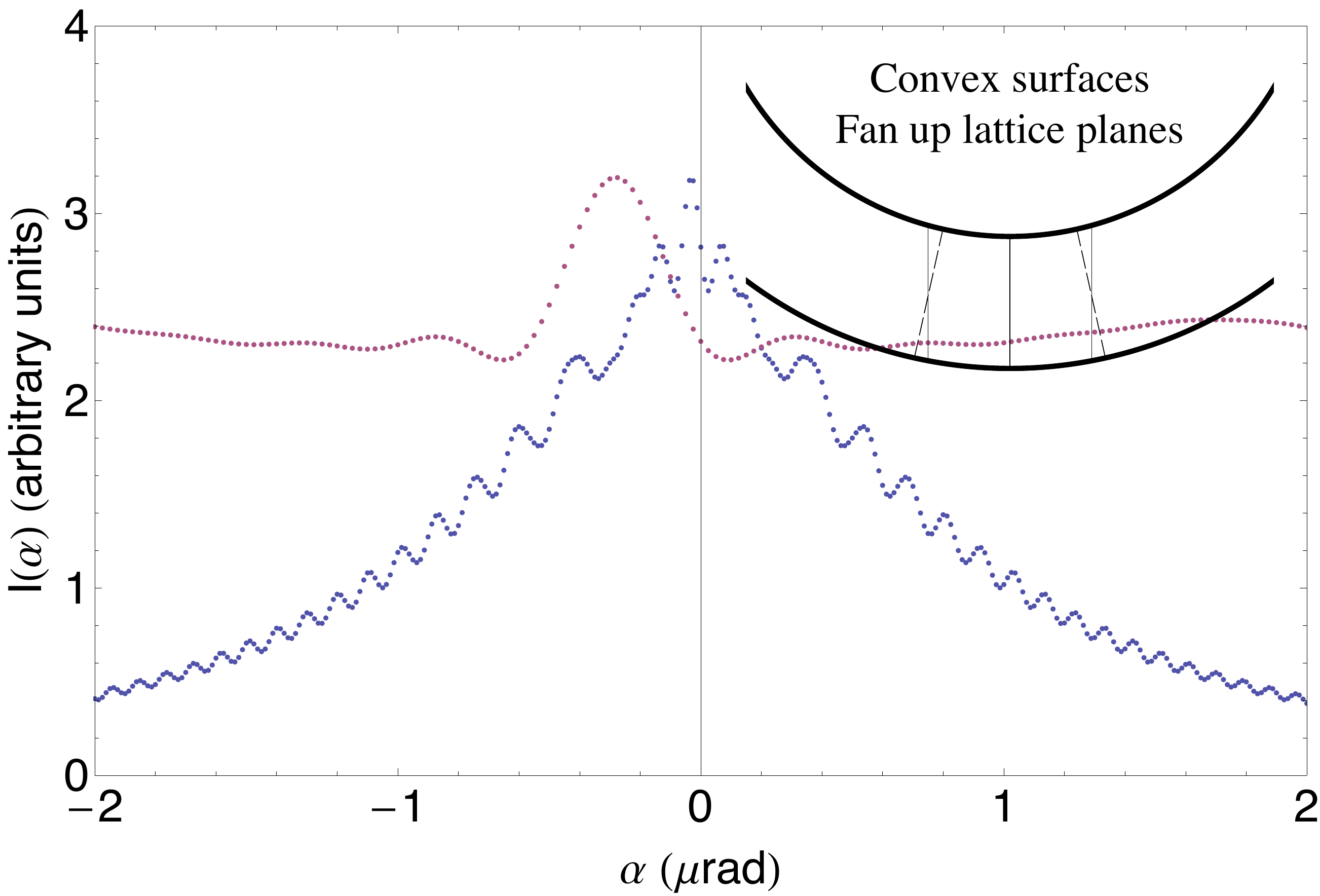}
\end{center}
\end{minipage}
\caption{
Non-dispersive (left) and dispersive (right)
rocking curves for a convex Si analyzer with a fan-down distortion.
Upper and lower curves refer to the energies of $17$ keV and $184$ keV, respectively.
Bragg planes are (220), the relevant parameter values are given in Table 1.
}
\label{convexd}
\end{figure}

Figs.~\ref{concavd} and \ref{convexd} show the rocking curves
for a concave and for a convex crystal,
in the non-dispersive and dispersive geometries.
Also in this case the peak shifts agree with the values obtained from equation (\ref{bra}).
By comparing the profiles in Figs.~\ref{pianaconcavd} and \ref{concavd} we observe a slight effect
depending on the surface curvature which, anyway, does not alter their symmetry;
furthermore, the wings of the upper curves do not appear in the range shown;
the same can be concluded from Figs.~\ref{pianaconvexd} and \ref{convexd}.

Additionally, the figures exemplify that peak shifts and,
consequently, $\Delta\alpha$, are independent of the surface geometry (flat or cylindrical,
concave or convex), but they depend only on the entrance-surface strain $\Delta d/d$.
This confirms that the flat-surface approximation used in Mana \emph{et al.}~(2004) was admissible.
To check further our numerical computations, we examined also two auxiliary cases,
when the Bragg planes are simply either contracted or expanded and when the
Bragg planes are not distorted. We have also carried out many numerical simulations
with different entrance-slit apertures; the above conclusion were always confirmed.

\section{Conclusions}
We have studied x- and $\gamma$-ray propagation in flat and cylindrically bent crystals.
We have used the relevant results to describe the operation of a Laue-Laue
diffractometer consisting of a flat collimating crystal and a bent analyzer crystal and we have
extended the results of a previous investigation of ours (Mana \emph{et al.,} 2004).
We have described the distortion charaterized by a constant strain gradient
in crystals having flat or cylindrical surfaces.

In both cases, in addition to numerical simulations, we have also given
exact solutions of the Takagi-Taupin equations
in the form of Riemann-Green integrals.
We have confirmed both analytical and numerical results that
the rocking curve shift does not depend on the shape of the
analyzer surface, but only on the lattice strain on the entrance surface.

Since the validity of these solutions -- via convolution integrals --
is not limited to flat and cylindrical surfaces,
we can extend such solutions either to the case when also the collimating crystal
is cylindrically bent, or to the case when the effect of the surface roughness
in x- and $\gamma$-ray diffractometry and interferometry is
not negligible.
Additionally, numerical simulations
open the way to a better understanding of the operation of bent-crystal
diffractometers, in particular when finite element solutions of the elasticity
equations are integrated into the Takagi-Taupin equations thus allowing us to characterize
the relevant lattice strains of the diffractometer crystals.

\ack
This work was supported by the {\it Museo Storico della Fisica e Centro
Studi e Ricerche ``Enrico Fermi'', Rome},
the {\it Regione Piemonte} and the {\it Compagnia San Paolo, Turin}. The
Institut Laue-Langevin is gratefully acknowledged for beam time allocation and support (M.~Jentschel)
in preliminary tests of our results.


\newpage
\begin{table}
\caption{\label{label}Parameters used for different energies}
\begin{tabular}{crrrrr}
ref.\   &$T$ (mm) &$L$ (m) &$w/2$  ($\mu$m) &$E$ (keV) &$R_0$ (m)\\
Mana     \emph{et al.} (2004) &0.5 &1  &50  &17  &160\\
Massa   \emph{et al.} (2006) &2.5 &16 &500 &184 &697\\
\end{tabular}
\end{table}

\clearpage

\clearpage

\clearpage

\clearpage

\end{document}